\documentclass[aps,prb,twocolumn,notitlepage,showpacs,superscriptaddress,10pt]{revtex4-2}%
\usepackage{graphicx}
\usepackage{amsmath}
\usepackage{bm}
\usepackage{amssymb}
\usepackage{color,soul}
\usepackage{amsfonts}%
\usepackage[caption=false]{subfig}
\usepackage{comment}
\usepackage{color,soul}
\usepackage[dvipsnames]{xcolor}
\definecolor{DarkGreen}{rgb}{0.0, 0.5, 0.0}
\usepackage[colorlinks,citecolor=DarkGreen,linkcolor=blue,linktocpage]{hyperref}
\usepackage[capitalize]{cleveref}
\usepackage{float}
\usepackage{tabularx}
\usepackage{array}   
\newcolumntype{L}{>{$}l<{$}} 
\newcolumntype{R}{>{$}r<{$}} 
\newcolumntype{C}{>{$}c<{$}} 
\usepackage{comment}
\usepackage[capitalize]{cleveref}
\usepackage{physics}

\usepackage{tikz}

\usepackage[utf8]{inputenc}

\newcommand{\tikzcircle}[2][red,fill=red]{\tikz[baseline=-0.5ex]\draw[#1,radius=#2] (0,0) circle ;}%

\begin{document}

\title{Edge-Edge Correlations without Edge-States: $\eta$-clustering State as  Ground State of the Extended Attractive SU(3) Hubbard Chain}

\author{Hironobu Yoshida}
\affiliation{Department of Physics, Graduate School of Science, The University of Tokyo, 7-3-1 Hongo, Tokyo 113-0033, Japan}
\author{Niclas Heinsdorf}
\affiliation{Max-Planck-Institut für Festkörperforschung, Heisenbergstrasse 1, D-70569 Stuttgart, Germany}
\affiliation{Department of Physics and Astronomy \& Stewart Blusson Quantum Matter Institute,
University of British Columbia, Vancouver BC, Canada V6T 1Z4}
\author{Hosho Katsura}
\affiliation{Department of Physics, Graduate School of Science, The University of Tokyo, 7-3-1 Hongo, Tokyo 113-0033, Japan}
\affiliation{Institute for Physics of Intelligence, The University of Tokyo, 7-3-1 Hongo, Tokyo 113-0033, Japan}
\affiliation{Trans-scale Quantum Science Institute, The University of Tokyo, 7-3-1, Hongo, Tokyo 113-0033, Japan}

\date{\today}

\begin{abstract}
We explore the phase diagram of the extended attractive SU($3$) Hubbard chain with two-body hopping and nearest-neighbor attraction at half-filling. In the large on-site attraction limit, we identify three different phases: phase separation (PS), Tomonaga-Luttinger liquid (TLL), and charge density wave (CDW). 
Our analysis reveals that the $\eta$-clustering state, a three-component generalization of the $\eta$-pairing state, becomes the ground state at the boundary between the PS and TLL phases. 
On an open chain, this state exhibits an edge-edge correlation, which we call boundary off-diagonal long-range order (bODLRO). Using the density matrix renormalization group (DMRG) method, we numerically study the phase diagram of the model with large but finite on-site interactions and find that the numerical results align with those obtained in the strong coupling limit. 
\end{abstract}

\maketitle

\section{Introduction}
In his seminal paper, C.N. Yang showed that the $\eta$-pairing states are exact eigenstates of the Hubbard model and have off-diagonal long-range order~\cite{yang_1989_eta} (ODLRO), which is a characteristic of superconductivity~\cite{yang_concept_1962}, crucial for the emergence of flux quantization and the Meissner effect~\cite{sewell_off-diagonal_1990, nieh_off-diagonal_1995}. Beyond their relevance for non-equilibrium superconductivity and superfluidity~\cite{kitamura_eta_2016,kaneko_photoinduced_2019,buvca_eta_2019,tindall_eta_2019,li_eta-pairing_2020,tsuji_tachyonic_2021,nakagawa_eta_2021},
$\eta$-pairing states enjoyed a renewed interest in recent years in the context of quantum many-body scars~\cite{scar1,scar2,scar3,scar4}. 
The $\eta$-pairing states are not proper scars of the Hubbard model, because they occupy distinct symmetry sectors and do not violate the eigenstate thermalization hypothesis~\cite{mark_eta_2020, moudgalya_2020_eta}. However, this can be rectified by modifying the Hamiltonian to turn them into true scar states~\cite{mark_eta_2020,moudgalya_2020_eta,pakrouski_many-body_2020,pakrouski_group_2021}.

\begin{figure}
    \centering  \includegraphics[width=\columnwidth]{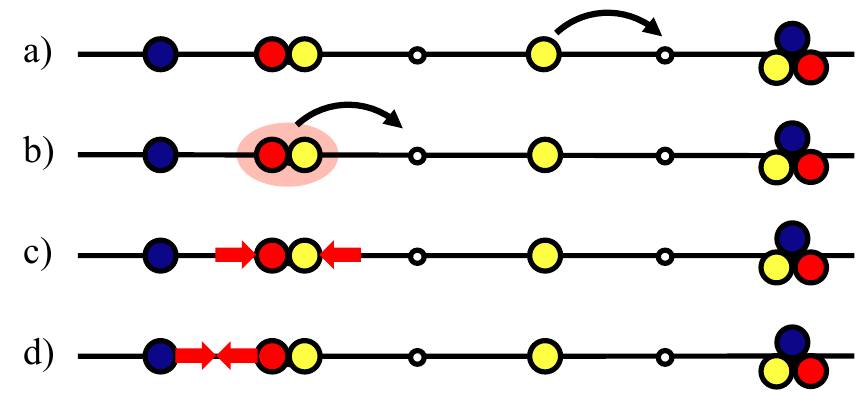}
    \caption{Schematic of the model. (a)~the one-body hopping term $\hat{H}_1$, (b)~the two-body hopping term $\hat{H}_2$, (c)~the on-site attractive interaction term $\hat{H}_U$, and (d)~the nearest-neighbor attractive interaction term $\hat{H}_V$.
    Blue, red, yellow (black, gray, white) balls represent fermions with flavors $\alpha=1,2,3$.}
    \label{fig:model}
\end{figure}

\begin{figure}
    \centering
    \includegraphics[width=\columnwidth]{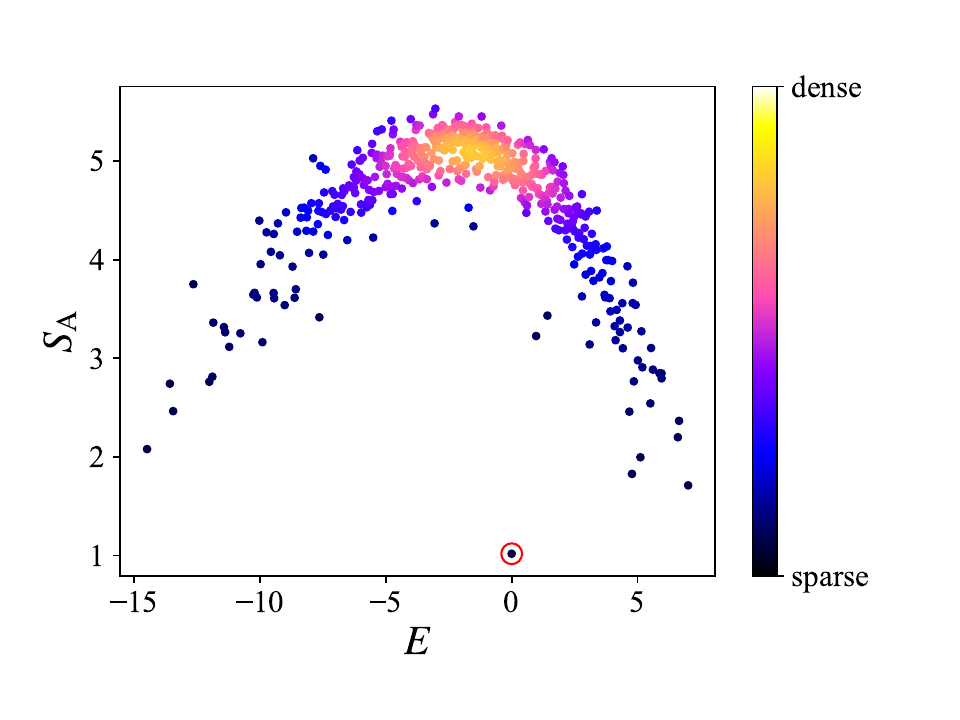}
    \caption{Half-chain bipartite entanglement entropy $S_A$ as a function of energy $E$ of the extended SU($3$) Hubbard chain \mbox{(see Fig.~\ref{fig:model})} with 
    $-t_1=t_2=1$, $U=-1$, $V=0$
    with $L=6$ sites at half-filling. The color scale for each dot indicates the density of data points.
    The $\eta$-clustering state is marked by a red circle. Since the model has SU($3$) and space inversion symmetries, we used the data in the SU($3$) singlet and the space inversion odd sector. The calculation was performed using QuSpin~\cite{weinberg_quspin1,weinberg_quspin2}.}
    \label{fig:scar}
\end{figure}

The $\eta$-pairing state can be generalized to higher-spin models~\cite{zhai2005two}, $N$-component fermionic models~\cite{yoshida_exact_2022,
nakagawa_exact_2022}, and multi-orbital models~\cite{imai_2024_systematic}. In this paper, we focus on a generalization to $N$-component fermions called the $\eta$-clustering states~\cite{yoshida_exact_2022}. 
These states can be made exact scar states by including $(N-1)$-body hoppings in the SU($N$) Hubbard model. See Fig.~\ref{fig:model} (b) for a schematic illustration of the two-body hopping. 
In Fig.~\ref{fig:scar}, we illustrate the position of the $\eta$-clustering state in the excitation spectrum of the model. Even though it is in the middle of the spectrum, it is well separated in entanglement entropy from other eigenstates. 

In this paper, we focus on one-dimensional systems, where the $\eta$-clustering states can be constructed for any integer $N$. Their properties vary depending on whether $N$ is even or odd: 
Even-$N$ clustering states have long-range ODLRO. In contrast, the correlations of an odd-$N$ clustering state decay exponentially. However, on an open chain, the edge-edge correlation persists in the thermodynamic limit~\cite{yoshida_exact_2022}. They can also be realized in the ground state of interacting and non-interacting Kitaev chains~\cite{1dinterac2,miao_exact_2017,miao_majorana_2018}, interacting fermionic~\cite{lang_topological_2015} and parafermionic chains~\cite{iemini_topological_2017}. We dub this type of order \textit{boundary off-diagonal long-range order} (bODLRO), and remark that this type of order does not necessarily imply the presence of localized edge modes.

Fermionic systems with $N$ internal degrees of freedom have successfully been realized in optical lattices~\cite{Abraham1997,Bartenstein2005,Fukuhara2007,Ottenstein2008,Huckans2009,
Taie2010,Taie2012,Desalvo2010,Lewenstein2012,Scazza2014,Zhang2014,Cazalilla2014,Pagano2014,Hofrichter2016}. In the limit of strong attractive interaction, one-dimensional fermionic cold atoms generically feature the molecular superfluid phases~\cite{oned2}, which are quasi-long-range ordered and formed from bound states made of $N$ fermions. For even $N$, a molecular superfluid phase in cold-atom setups~\cite{oned1,oned2,oned3,oned4,marti_charge_2024} is also associated with higher-charge superconductors in solid-state setup~\cite{herland_phase_2010,4ekivelson,4efernandes,magnonSC}.
For $N=3$, the formation of three-body bound states has been discussed in the context of Efimov states~\cite{efimov_weakly_1970,efimov_energy_1970,naidon_efimov_2017} and Cooper triples~\cite{niemann_pauli_2012,tajima_cooper_2020,akagami_condensation_2021,tajima_three_2021} in continuous systems.

Since it is generally difficult to prepare specific high-energy eigenstates experimentally, one way to achieve $\eta$-clustering states is to realize a model that has them as its ground state. In Ref.~\cite{yoshida_exact_2022}, it was shown that $\eta$-clustering states are the unique ground state of a suitably tailored Hamiltonian. However, it is required that the one- and two-body hopping amplitudes be equal, which is highly demanding experimentally in a cold-atom setup. In this study, we explore the phase diagram of the extended attractive SU($3$) Hubbard chain with a two-body hopping and nearest-neighbor interaction (See Fig.~\ref{fig:model}). 

In previous work~\cite{perez-romero_phase_2021}, the extended SU(3) Hubbard chain with nearest-neighbor interaction at one-third filling was studied numerically. For attractive on-site interactions, the phase diagram consists of four distinct phases: phase separation (PS), charge density wave (CDW), pairing phase, and an intermediate phase. In this work, we further introduce two-body hopping terms~\cite{penson_real-space_1986,affleck_field-theory_1988, hui_penson_1993,arrachea_ground-state_1997, japaridze_eta_2001,dutta_non-standard_2015,sous_light_2018,moller_magnon_2019,sous_fracton_2020,gotta_two-fluid_2021} and 
focus on the half-filling sector. 
When the on-site interaction term is sufficiently large, we can treat these terms as perturbations. By mapping the system to a spin-$1/2$ XXZ chain, we show that there are three phases: 
PS, Tomonaga-Luttinger liquid (TLL), and CDW. 
The $\eta$-clustering state arises as a gapless ground state along the phase boundary between the PS and TLL phases.
Notably, it becomes the ground state when the two-body hopping amplitude and nearest-neighbor interaction are small compared to the one-body hopping amplitude. This is in contrast to the attractive SU($3$) Hubbard chain without two-body hopping and nearest-neighbor interaction, where the ground state only shows the CDW phase at half-filling~\cite{zhao_insulating_2007,nonne_competing_2011,capponi_2016}.

Using the density matrix renormalization group (DMRG), we numerically explore the phase diagram. We confirm that results are consistent with the perturbative result even for finite but large on-site interactions. We compute the entanglement entropy of our $\eta$-clustering ground state, showing excellent agreement with analytic predictions, and further demonstrate that it exhibits bODLRO. Large parts of our phase diagram are separated by a Berezinskii–Kosterlitz–Thouless (BKT) transition, which is numerically challenging to pinpoint, because the crossover from exponentially to algebraically decaying correlation functions is hard to capture for numerically tractable system sizes. Instead of determining the exact shape of the boundary between the TLL and CDW phases, we focus on several paths in the phase diagram and determine the BKT transition points along these paths by detecting level crossings due to the emergent SU(2) symmetry.

The article is organized as follows. In Sec.~\ref{sec:model}, we introduce the extended SU($3$) Hubbard model and $\eta$-clustering states, and review their basic properties. 
In Sec.~\ref{sec:perturbation}, we derive the effective Hamiltonian in the limit of strong on-site attraction and discuss the ground-state phase diagram.
In Sec.~\ref{sec:numerics}, we present numerical results for large but finite on-site interactions. By examining specific paths on the phase diagram and detecting phase transitions, we validate the effective Hamiltonian. Finally, we discuss our results in Sec.~\ref{sec:discussion}.

\section{Extended attractive SU($3$) Hubbard Model}\label{sec:model}
\subsection{The Hamiltonian}

We consider a chain of three-component fermions with $L$ lattice sites. For each site $j=1,\ldots, L$, we denote by $\hat{c}^\dagger_{j,\alpha}$ and $\hat{c}_{j,\alpha}$ the creation and annihilation operators, respectively, of a fermion with flavor $\alpha=1,2,3$. We write the normalized vacuum state annihilated by all $\hat{c}_{j,\alpha}$ as $\ket{0}$. The whole Fock space $\mathcal{V}$ is spanned by states of the form $\{ \prod_{j=1}^L \prod_{\alpha=1}^3 (\hat{c}^\dagger_{j,\alpha})^{n_{j,\alpha}} \} \ket{0}$ $(n_{j,\alpha}=0,1)$. The number operators are defined as $\hat{n}_{j,\alpha} =\hat{c}^\dagger_{j,\alpha} \hat{c}_{j,\alpha}$ and $\hat{n}_j=\sum_{\alpha=1}^{3} \hat{n}_{j,\alpha}$.
We define three-particle creation operators
$\hat{\eta}^\dagger_j = \hat{c}^\dagger_{j,1}\hat{c}^\dagger_{j,2}\hat{c}^\dagger_{j,3}$ and two-particle creation operators $\hat{\overline{c}}^\dagger_{j,1}= \hat{c}^\dagger_{j,2}\hat{c}^\dagger_{j,3}$, $\hat{\overline{c}}^\dagger_{j,2}= \hat{c}^\dagger_{j,3}\hat{c}^\dagger_{j,1}$, and $\hat{\overline{c}}^\dagger_{j,3}= \hat{c}^\dagger_{j,1}\hat{c}^\dagger_{j,2}$.

Let us consider the Hamiltonian of the extended SU($3$) Hubbard model with open boundary conditions,
\begin{align}\label{eq:hamiltonian}
    \hat{H} &=\hat{H}_{1}+\hat{H}_{2}+\hat{H}_{U}+\hat{H}_{V}, \\ \hat{H}_{1} &=t_1 \sum_{j=1}^{L-1} \sum_{\alpha=1}^{3}\left(\hat{c}_{j, \alpha}^{\dagger} \hat{c}_{j+1, \alpha}+\text { h.c. }\right), \\ \hat{H}_{2} &=-t_2 \sum_{j=1}^{L-1} \sum_{\alpha=1}^{3}\left(\hat{\bar{c}}_{j, \alpha}^{\dagger} \hat{\bar{c}}_{j+1, \alpha}+\text { h.c. }\right), \\
    \hat{H}_{U} &=\frac{U}{2} \sum_{j=1}^{L} \hat{n}_{j}\left(\hat{n}_{j}-3\right),\\
    \hat{H}_{V} &=V \sum_{j=1}^{L-1} \left(\hat{n}_{j}-\frac{3}{2}\right)\left(\hat{n}_{j+1}-\frac{3}{2}\right).
\end{align}
A schematic of each term in the Hamiltonian is shown in Fig.~\ref{fig:model}. 
The first term $\hat{H}_1$ describes the one-body hopping term with amplitude $t_1 \in \mathbb{R}$, and the second term $\hat{H}_{2}$ represents the two-body hopping term with amplitude $t_2 \in \mathbb{R}$. \footnote{Without loss of generality, we can assume that $t_1 \geq 0$; if $t_1 < 0$, we can change its sign by flipping the sign of $\hat {c}_{j,\alpha}$ at every other site, i.e., $\hat{c}_{j,\alpha} \to (-1)^j \hat{c}_{j,\alpha}$. Note that this transformation does not change the signs of the other parameters of the model.} The third term $\hat{H}_U$ and the final term $\hat{H}_V$ represent on-site and next-nearest attractive ($U<0$, $V<0$) interactions with a potential uniform except at the boundaries~\footnote{Note that these terms can be rewritten as \\$\hat{H}_{U} =U \sum_{j=1}^{L} \sum_{1\leq\alpha<\beta\leq3}\left(\hat{n}_{j,\alpha}-\frac{1}{2}\right)\left(\hat{n}_{j,\beta}-\frac{1}{2}\right)+\text{const.}$ and $\hat{H}_{V} =V \sum_{j=1}^{L-1} \sum_{\alpha,\beta = 1}^3 \left(\hat{n}_{j,\alpha}-\frac{1}{2}\right)\left(\hat{n}_{j+1,\beta}-\frac{1}{2}\right)$.}.
This model can be regarded as an SU($3$) generalization of the extended Hubbard model discussed in Ref.~\cite{dutta_non-standard_2015}.

\subsection{Symmetries of the Hamiltonian}
Let us first discuss the symmetries of the Hamiltonian. First, the Hamiltonian has the U(3)= U(1)$\times$ SU($3$) symmetry. To see this, we define the operators $\hat{F}^{\alpha, \beta}= \sum_{j=1}^L \hat{c}^\dagger_{j,\alpha}\hat{c}_{j,\beta}$. Here, $\hat{F}^{\alpha, \alpha}$ is the total number operator of fermions with flavor $\alpha$, while $\hat{F}^{\alpha, \beta}$ ($\alpha \ne \beta$) are flavor-raising and lowering operators. Then, all $\hat{F}^{\alpha,\beta}$ operators commute with each of $\hat{H}_{1}$, $\hat{H}_{2}$, $\hat{H}_{U}$, and $\hat{H}_{V}$.
From the operators $\hat{F}^{\alpha,\beta}$, one can construct new operators as $\hat{F}=\sum_{\alpha=1}^3 \hat{F}^{\alpha, \alpha}$ and $\hat{T}^a = \sum_{\alpha,\beta=1}^3 \mathcal{T}^a_{\alpha,\beta}\hat{F}^{\alpha, \beta}$ for $a=1,\ldots,8$, where $\mathcal{T}^a_{\alpha,\beta}$ are 
the Gell-Mann matrices. Thus it is concluded that the Hamiltonian has a global U(3)= U(1)$\times$ SU(3) symmetry. Due to the U($1$) symmetry, we can fix the total fermion number $N_{\mathrm{tot}}$. Then, the filling factor is defined as $\rho=\frac{N_{\mathrm{tot}}}{3L}$. When $F$ is a multiple of 3, a state $\ket{\Psi_\text{sing}}$ is an SU(3) singlet if $\hat{T}^a \ket{\Psi_\text{sing}}=0$ for all $a$. The three-particle operators $\hat{\eta}^\dagger_j$ ($\hat{\eta}_j$) commute with $\hat{T}^a$, and they create (annihilate) a local SU(3) singlet at site $j$.

\subsection{$\eta$-Clustering State}
In this section, we define the $\eta$-clustering states and review their properties~\cite{yoshida_exact_2022}. To define $\eta$-clustering states, let us define $\eta$-operators as
\begin{equation}    \hat{\eta}=\sum_{j=1}^L (-1)^j \hat{U}_{j-1} \hat{\eta}_{j},\ 
\hat{\eta}^\dagger=\sum_{j=1}^L (-1)^j \hat{U}_{j-1} \hat{\eta}^{\dagger}_{j},
\label{eq:def_eta_op}
\end{equation}
where 
\begin{equation}\label{eq:string_unitary}
\hat{U}_{j-1}=\left\{
\begin{array}{cc}
1 & {\rm if}~j=1 \\
\exp \left( i \pi \sum_{k=1}^{j-1} \hat{n}_k\right) & {\rm if}~j>1
\end{array}
\right.
\end{equation}
is a Hermitian and unitary operator~\footnote{This $\eta$-operator 
generalizes the one introduced in Ref.~\cite{yang_1989_eta}. In 
the SU(2) Hubbard model, the $\eta$-operator is defined as $\hat{\tilde{\eta}}_2^\dagger=\sum_{j=1}^L (-1)^j\hat{c}^{\dagger}_{j,1}\hat{c}^{\dagger}_{j,2}$. One might naively attempt to generalize it to the SU(3) case 
by defining $\hat{\tilde{\eta}}_3^\dagger=\sum_{j=1}^L (-1)^j \hat{\eta}_{j}^\dagger$, but this approach fails because $\hat{\tilde{\eta}}_3^\dagger$ squares to zero. To circumvent this issue, we introduced the Jordan-Wigner string $\hat{U}_{j-1}$ so that $({\hat \eta}^\dagger)^2$ remains nonzero.}.
By repeatedly applying $\hat{\eta}^\dagger$ to the vacuum state, we have $\eta$-clustering states with fermion number $N_{\mathrm{tot}}=3M$:
\begin{equation}
    \ket{\Phi^{L}_{M}}
    =\frac{1}{M!}(\hat{\eta}^\dagger)^M \ket{0}  \quad (M=0,\ldots, L).
\label{eq:def_eta_state}
\end{equation}

\begin{figure}
    \centering
    \includegraphics[width=0.75\linewidth]{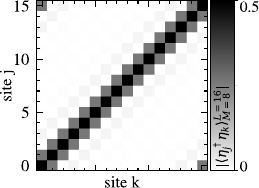}
    \caption{The absolute value of the singlet correlation function $\langle \hat{\eta}^\dagger_j \hat{\eta}_k \rangle^{L=16}_{M=8}$ calculated from Eq.~\eqref{eq:singlet_correlation}. A darker color indicates a larger correlation. The correlation decays exponentially in the bulk, but the end-to-end correlation survives as shown in the plot's upper left and lower right.}
    \label{fig:eta_correlation}
\end{figure}

Let us review their properties. 
First, when $t_1=-t_2$ and $V=0$, one can check that $\ket{\Phi^{L}_{M}}$ is a (high-energy) eigenstate of $\hat{H}$ (See Fig.~\ref{fig:scar}). Secondly, their entanglement entropy obeys a sub-volume law. This can be seen by computing the entanglement entropy across a middle cut. 
We partition the $L$ sites $j=1,\ldots,L$ into a subsystem $A$ $(j=1,\ldots, L_A)$ and a subsystem $B$ $(j=L_A+1,\ldots, L)$. The entanglement entropy $S_A$ of the subsystem $A$ 
was obtained previously~\cite{vafek_entanglement_2017,vedral_high-temperature_2004,fan_entanglement_2005}:
\begin{align}
    S_A&=-\sum_{m=0}^{L_A}\lambda_m \log \lambda_m,
    \label{eq:ent_entropy1}\\
    \lambda_m &= \binom{L_A}{m}\binom{L-L_A}{M-m} \left/ \binom{L}{M}\right..
    \label{eq:ent_entropy2}
\end{align}
The central-cut entanglement entropy for even $L$ can be obtained by substituting $L_A=L/2$ to Eqs. \eqref{eq:ent_entropy1} and \eqref{eq:ent_entropy2}. For large $L$ with fixed $\rho=\frac{M}{L}$, the leading term in $S_A$ is logarithmic in $L$, proving that the $\eta$-clustering states have sub-volume law entanglement entropy.
Finally, 
the end-to-end singlet correlation in these states exhibits anomalous behavior. The singlet correlation function in a normalized state $\ket{\psi}$ is defined as $\langle \hat{\eta}^\dagger_j \hat{\eta}_k \rangle=\bra{\psi} \hat{\eta}^\dagger_j \hat{\eta}_k \ket{\psi}$.
Then, for an $\eta$-clustering state $\ket{\Phi^{L}_{M}}$, 
it was shown in~\cite{yoshida_exact_2022} that 
\begin{align}
    \langle \hat{\eta}^\dagger_j \hat{\eta}_k \rangle^L_M=\dfrac{\sum_{l=l_{\mathrm{min}}}^{l_{\mathrm{max}}} (-1)^{l}\binom{L-r-1}{j} \binom{r-1}{M-l-1}}{(-1)^{M+r-1} \binom{L}{M}},
\label{eq:singlet_correlation}
\end{align}
where $r=|j-k|$, $l_{\mathrm{min}}=\operatorname{max}\{0,M-r\}$, and ${l_{\mathrm{max}}}={\operatorname{min}\{L-r-1,M-1\}}$. Fig.~\ref{fig:eta_correlation} shows the singlet correlation function in the case $L=16$, $M=8$. Unlike the $\eta$-pairing state in the SU($2$) Hubbard model, the absolute value of this correlation function decays exponentially with distance $r\geq1$ as $\rho (1-\rho)|1-2\rho|^{r-1}$ in the bulk in the thermodynamic limit $L\to \infty$~\cite{yoshida_exact_2022}. However, the end-to-end correlation 
persists in the thermodynamic limit, 
a feature we call boundary off-diagonal long-range order (bODLRO). At half-filling ($\rho=1/2$), there is no correlation when $r>1$ in the limit $L\to \infty$. However, when $L$ is finite, the correlation does not vanish even when $r>1$ due to the finite-size effect as shown in Fig.~\ref{fig:eta_correlation}.

\section{\label{sec:perturbation}Effective Hamiltonian at large negative $U$}
While the model \eqref{eq:hamiltonian} is not tractable analytically for a generic set of parameters, much insight can be gained from the analysis in the strong coupling limit.
In the limit $|U|\gg |t_1|, |t_2|, |V|$, the effective Hamiltonian is described by a spin-$1/2$ chain, where the up and down spins at each site correspond to a singlet and an empty state, respectively~\cite{efetov_correlation_1975,emery_theory_1976}. We treat $\hat{H}_U$ as the unperturbed Hamiltonian and $\hat{H}_1$, $\hat{H}_2$, $\hat{H}_V$ as perturbations. We first examine the ground states of $\hat{H}_U$. Since each term in $\hat{H}_U$ is minimized when the occupation number is $0$ or $3$ for all $j$, the ground states of $\hat{H}_U$ are spanned by states  
of the form $\prod_{j=1}^L  (\hat{\eta}^\dagger_{j})^{n_{j}}  \ket{0}$ $(n_j=0,1)$.  
In the following, we write the space spanned by these states as $\mathcal{W}$ and denote the orthogonal projection to $\mathcal{W}$ as $\hat{P}$.

In the subspace $\mathcal{W}$, the states can be mapped to  
those of spin-$1/2$ chains by  
identifying $\ket{0}$  
with the all-down state $\ket{\Downarrow}=\bigotimes_{j=1}^L \ket{\downarrow}_j$ and defining spin-$1/2$ operators as
\begin{align}
     \hat{S}_j^+ &=(-1)^j\hat{P}\hat{U}_{j-1}\hat{\eta}_{j}^{\dagger} \hat{P}, \label{eq:eta_spin1}\\
     \hat{S}_j^- &=(-1)^j\hat{P} \hat{\eta}_j \hat{U}_{j-1}^\dagger\hat{P}, \label{eq:eta_spin2}\\
     \hat{S}_j^z &=\hat{P} \left(\hat{\eta}_{j}^{\dagger} \hat{\eta}_j-\frac{1}{2}\right) \hat{P},
     \label{eq:eta_spin3}
\end{align}
where $\hat{U}_{j-1}$ is defined in Eq.~\eqref{eq:string_unitary}. 
These operators satisfy the commutation relations $\left[\hat{S}^+_j,\hat{S}^-_k\right]=2\delta_{j,k} \hat{S}^z_j$ and $\left[\hat{S}^z_j,\hat{S}^\pm_k\right]=\pm \delta_{j,k} \hat{S}^\pm_j$. Then, the total spin operators are written in terms of $\eta$-operators:
\begin{align}  
    \hat{S}_\mathrm{tot}^+&=\sum_{j=1}^L\hat{S}_j^+=\hat{P}\hat{\eta}^\dagger\hat{P}, \\
    \hat{S}_\mathrm{tot}^-&=\sum_{j=1}^L\hat{S}_j^-=\hat{P}\hat{\eta}\hat{P}, \\
    \hat{S}_\mathrm{tot}^z&=\sum_{j=1}^L\hat{S}_j^z=\hat{P} \sum_{j=1}^L\left(\hat{\eta}_{j}^{\dagger} \hat{\eta}_j-\frac{1}{2}\right) \hat{P}.
\label{eq:total_spin}
\end{align}
Note that they satisfy $\left[\hat{S}^+_\mathrm{tot},\hat{S}^-_\mathrm{tot}\right]=2\hat{S}^z_\mathrm{tot}$ and $\left[\hat{S}^z_\mathrm{tot},\hat{S}^\pm_\mathrm{tot}\right]= \hat{S}^\pm_\mathrm{tot}$.
Let us see the effect of the perturbations $\hat{H}_1$, $\hat{H}_2$ and $\hat{H}_V$ up to the second order. In Appendix \ref{sec:appendix_perturbation}, we 
show that the effect of the perturbation is described by the XXZ Hamiltonian (up to a constant energy shift)
\begin{equation}
    \hat{H}_{\mathrm{eff}}=\sum_{j=1}^{L-1} \left[\frac{J_x}{2}\left(\hat{S}_j^+ \hat{S}_{j+1}^-+\hat{S}_j^- \hat{S}_{j+1}^+\right)+J_z\hat{S}_j^z \hat{S}_{j+1}^z\right],
    \label{eq:pert_xxz}
\end{equation}
where 
\begin{align}
    J_x= 6\frac{t_1 t_2}{U} \text{ and } J_z=-3\frac{t_1^2+t_2^2}{U}+9V.
    \label{eq:pert_jxjz}
\end{align}

Note that by applying a $\pi$ rotation about the $z$-axis on the odd sites, one can flip the sign of $J_x$. 
Therefore, the phase diagram does not depend on the sign of $J_x$, and from this, we can also see that the phase diagram is independent of the sign of $t_2/t_1$. From now on, we focus on the half-filling sector, i.e., $\rho=1/2$ and thus $S^z_\mathrm{tot}=0$. At this filling, the ground-state phase diagram of the XXZ Hamiltonian~\eqref{eq:pert_xxz} exhibits three distinct phases~\cite{giamarchi_quantum_2003}:
\begin{enumerate}
    \item Ising ferromagnet with a domain wall in the middle if $J_z<-|J_x|$,
    \item Tomonaga-Luttinger liquid (TLL) if $-|J_x|<|J_z|\leq|J_x|$,
    \item Ising antiferromagnet if $|J_x|<J_z$.
\end{enumerate}
At the boundary between the Ising ferromagnetic and TLL phases ($J_z=-|J_x|$), the ground state is 
a ferromagnet of the form $(\hat{S}_\mathrm{tot}^+)^{L/2} \ket{\Downarrow}$. From the TLL into the Ising antiferromagnetic phase or vice versa, the transition is of BKT type.

\begin{figure}
    \centering
    \includegraphics[width=\linewidth]{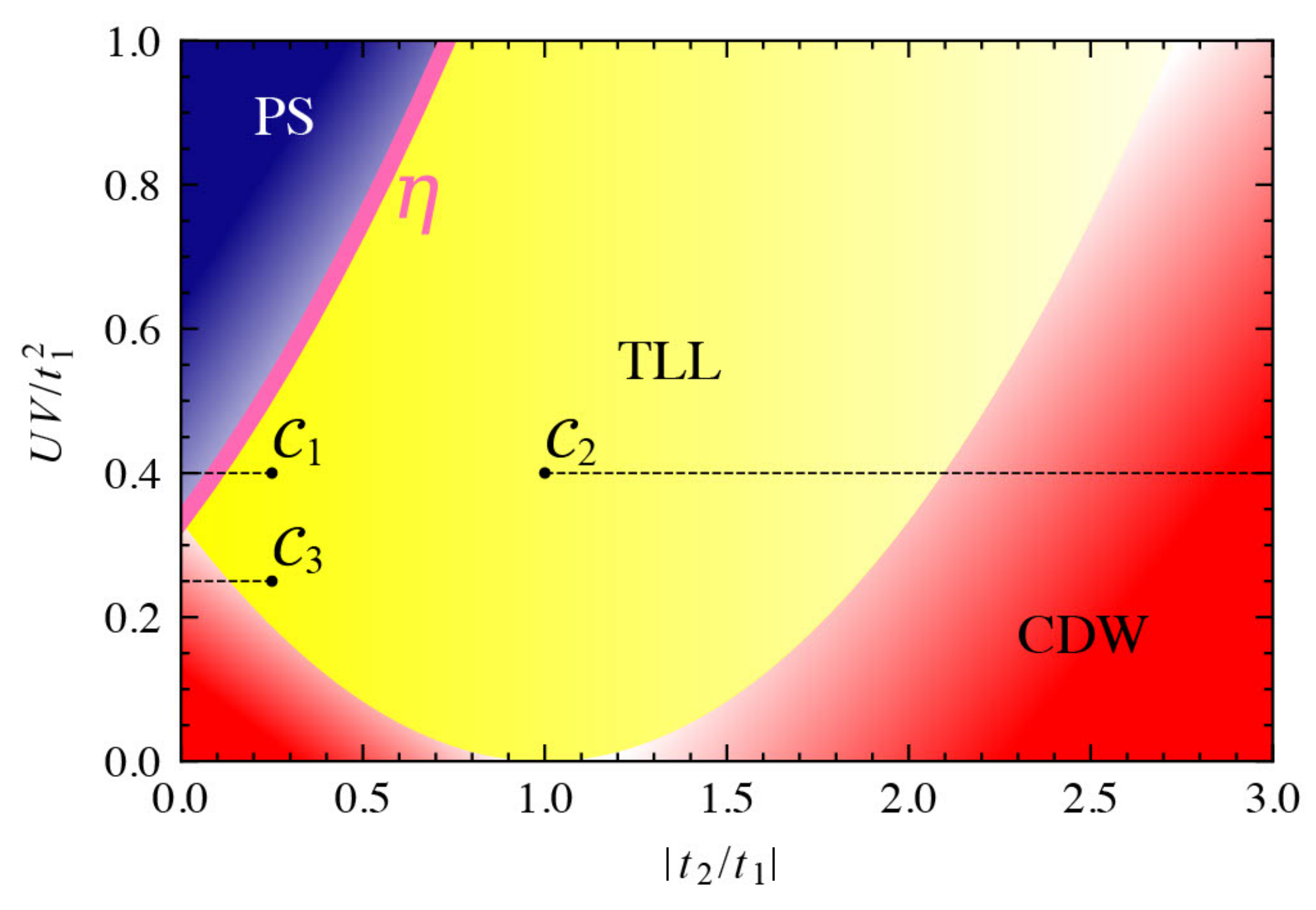}
    \caption{Phase diagram of the extended attractive SU(3) Hubbard chain with strong on-site attractive interaction at half-filling, obtained by perturbation theory. There are three extended regions: phase separation (PS) in blue, Tomonaga-Luttinger Liquid (TLL) in yellow, and charge density wave (CDW) in red. At the transition between the PS and TLL state, the ground state of the system is the $\eta$-clustering state. From the TLL into the CDW phase or vice versa, the transition is of BKT type. Note that the phase diagram is independent of the sign of $t_2/t_1$. We present numerical results from three paths through this diagram along $\mathcal{C}_{1,2,3}$.}
    \label{fig:phasediagram}
\end{figure}

In the language of fermions, the above phases translate into the following phases:
\begin{enumerate}
    \item Phase separation (PS)~\footnote{
    Here, phase separation refers to a state in which particles tend to aggregate due to attractive interactions, resulting in two distinct regions: one particle-rich and the other particle-poor.} if $J_z<-|J_x|$,
    \item TLL if $-|J_x|<|J_z|\leq|J_x|$,
    \item Charge density wave (CDW) if $|J_x|<J_z$.
\end{enumerate}
At $J_z=-|J_x|$, the $\eta$-clustering state $(\hat{\eta}^\dagger)^{L/2} \ket{0}$ is the ground state,
and the BKT-type transition occurs at the boundary between the TLL and CDW phases~\footnote{This ground state is stable against certain types of perturbations. For example, the $\eta$-clustering state is a ground state of local Hamiltonians 
$\hat{h}_{j,j+1} = \frac{1}{2}(\hat{\eta}^{\dagger}_{j}\hat{\eta}_{j+1}+\text{h.c.)}
-\frac{1}{9} (\hat{n}_j-\frac{3}{2}) (\hat{n}_{j+1}-\frac{3}{2})$~\cite{yoshida_exact_2022}. Thus, even if we add $\hat{h}_{j,j+1}$ terms to our Hamiltonian, $\eta$-clustering state 
is expected to remain a ground state. The first term of $\hat{h}_{j,j+1}$ describes the three-body hopping term. The second term represents the next-nearest attractive interactions, which can be absorbed into $\hat{H}_V$.}. By substituting \eqref{eq:pert_jxjz} into the above conditions, we obtain the phase diagram (Fig.~\ref{fig:phasediagram}). We emphasize that the $\eta$-clustering state becomes the ground state when the two-body hopping amplitude and nearest-neighbor interaction are small compared to the one-body hopping amplitude. For example, when $|U/t_1|=10$, it becomes the ground state at $|t_2/t_1|=0.040$ and $|V/t_1|=0.036$.

As we have seen, the model \eqref{eq:hamiltonian} simplifies drastically in the large-$|U|$ limit. Another tractable case is the limit where $t_1 = 0$, i.e., the model without one-body hopping. Interestingly, in this limit, the model exhibits Hilbert space fragmentation~\cite{sala_ergodicity_2020,khemani_localization_2020,zadnik_folded_2021,pozsgay2021integrable,moudgalya_thermalization_2021, scar1}. 
In Appendix~\ref{sec:fragmentation}, we focus on the subspace spanned by states without singly or triply occupied sites and show that the model within this subspace is equivalent to Maassarani’s XXC model~\cite{maassarani_xxc_1998, arnaudon1998integrable}, a typical example of an integrable model exhibiting Hilbert space fragmentation. We further argue that the effective Hamiltonian within each fragmented sector can be exactly mapped to the spin-$1/2$ XXZ Hamiltonian, which means that the model at $t_1=0$ is partially integrable~\cite{zhang2022hidden, Matsui_2024_1, matsui2024boundary}.

\section{Numerical Results}\label{sec:numerics}

To verify the predictions of the perturbation 
analysis, we numerically calculate the entanglement entropy, the density-density correlation, and the singlet correlation in the ground state using the Density Matrix Renormalization Group (DMRG) method. The details of the method are presented in Appendix~\ref{app:dmrg}. 
To detect the phase transition, we present numerical results from three paths along $\mathcal{C}_{1,2,3}$ in Fig.~\ref{fig:phasediagram}. 
We fix $t_1 = 1$ and $U = -10t_1$ so that along path $\mathcal{C}_{1}$ and $\mathcal{C}_{3}$ both $t_2$ and $V$ can be thought of as small. The system size is fixed to $L=16$ unless otherwise noted.

\subsection{$\eta$-clustering state at the boundary between the PS and TLL phases}

\begin{figure}
    \centering
    \includegraphics{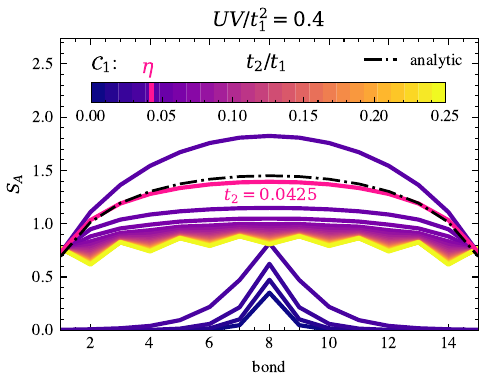}
    \caption{Entanglement entropies of bipartitions at different bonds for values of $t_2/t_1$ along $\mathcal{C}_1$. In the PS region there is only entanglement between the occupied and unoccupied halves of the chain, whereas the TLL states show a more flat, but weakly oscillating behavior that stems from the competition with the CDW phase. At intermediate $t_2/t_1$ we see the convex shape that is expected for $\eta$-clustering states. The entanglement entropy of the $\eta$-clustering state calculated from Eqs.~\eqref{eq:ent_entropy1} and \eqref{eq:ent_entropy2} is shown in dash-dotted line.}
    \label{fig:EE_C1}
\end{figure}

Figure~\ref{fig:EE_C1} shows the entanglement entropies of bipartitions at different bonds for values of $t_2/t_1$ along $\mathcal{C}_1$. When $t_2/t_1$ is small, the entanglement is very small except for the central cut, which is consistent with the PS phase. By increasing $t_2/t_1$, the entanglement entropy is maximized at $t_2/t_1=0.04$ and 
we can see the convex shape, which is expected for the $\eta$-clustering states and consistent with its gaplessness. Finally, we observe a weakly oscillating behavior, reminiscent of that of a ($1+1$)-dimensional conformal field theory with open boundaries~\cite{laflorencie2006boundary, demidio_2015,kim_2016}. 

\begin{figure}
    \centering
    \includegraphics{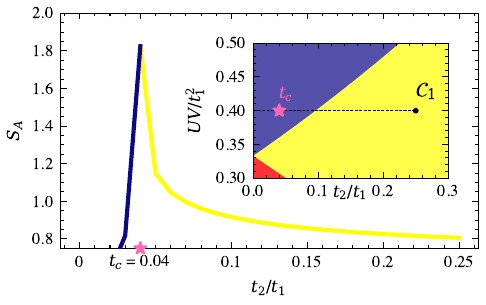}
    \caption{The central cut entanglement entropy $S_A$ as a function of $t_2/t_1$ along $\mathcal{C}_1$ with a sharp peak at $t_c = 0.04$ going from the PS (blue) into the TLL (yellow) phase while passing through the $\eta$-clustering ground state. The pink star marks the transition point in the phase diagram shown in the inset.}
    \label{fig:central_charge}
\end{figure}

The phase transition from the PS phase to the TLL phase can be detected by the central cut entanglement entropy $S_A$ (Fig.~\ref{fig:central_charge}). By increasing $t_2/t_1$ along $\mathcal{C}_1$, $S_A$ increases and we can see a sharp peak at the phase transition point $t_2/t_1 = 0.04$. 
To identify the nature of the TLL phase, we study the scaling of $S_A$ with the system size.
In a one-dimensional critical system with system size $L$, whose continuum limit is a conformal field theory with central charge $c$, $S_A$ behaves as~\cite{calabrese_entanglement_2004} 
\begin{equation}
    S_A=\frac{c}{6}  \log L+\text{const.}
\end{equation}
At $t_2/t_1=0.2$, which is well inside the TLL phase, we plotted $S_A$ as a function of 
$\log L$ (with $L=36$, 40, 44, 48, 52) and fitted linearly to compute a central charge of $c\approx 1.052$, close to the expected value for a TLL of $c=1$.

\begin{figure}
    \centering
    \includegraphics{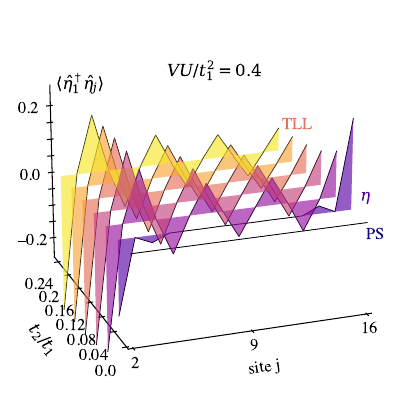}
    \caption{Singlet correlation $\langle \hat{\eta}^\dagger_i \hat{\eta}_j  \rangle$ as a function of distance with one site fixed to the boundary $i = 1$ for different values of $t_2/t_1$ along $\mathcal{C}_1$. In the TLL region (yellow - orange) the correlation decays, while in the PS state (blue) it is completely flat. The $\eta$-clustering state (purple) at $t_2/t_1 = 0.04$ has sharp peaks at the boundaries of the chain indicating bODLRO. 
    }
    \label{fig:singlet_correlation}
\end{figure}

Finally, Fig.~\ref{fig:singlet_correlation} shows the singlet correlation function as a function of distance. The sharp peak at the boundaries of the chain at $t_2/t_1 = 0.04$ is another piece of evidence that the $\eta$-clustering state lies on 
the boundary of the PS and TLL phase.

\subsection{BKT transition}

\begin{figure}
    \centering
    \includegraphics{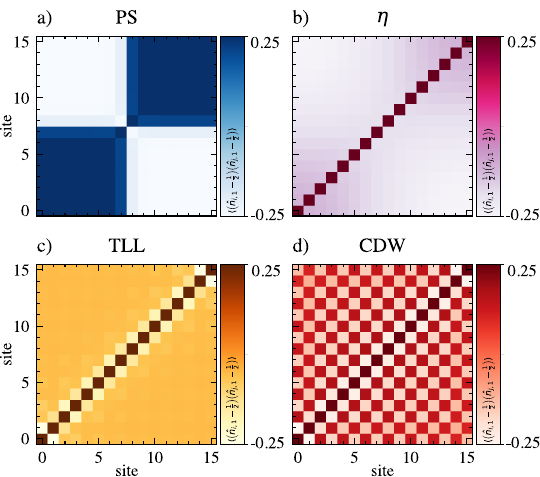}
    \caption{Density-Density correlation functions $\langle \left(\hat{n}_{i,1} - \frac{1}{2} \right) \left(\hat{n}_{j,1} - \frac{1}{2} \right) \rangle$ for 
    (a) a PS ground state with $VU/t_1^2 = 0.4$ and $t_2/t_1=0$, (b) the $\eta$-clustering state with $VU/t_1^2 = 0.4$ and $t_2/t_1=0.04$, (c) a ground state from the TLL region with $VU/t_1^2 = 0.4$ and $t_2/t_1=0.25$ and (d) a CDW ground state with $VU/t_1^2 = 0.25$ and $t_2/t_1=0$. 
    The PS and the CDW ground state can clearly be diagnosed by this correlation function. The $\eta$-clustering and TLL-state both show decaying correlations. Because 
    $|U| \gg t_1, t_2, V$, the other fermion flavors $\alpha = 2,3$ show the same behavior. }
    \label{fig:dens_dens}
\end{figure}

Figure~\ref{fig:dens_dens} shows the density-density correlation functions in each phase in the
phase diagram. This correlation function can diagnose the PS and the CDW ground state. The $\eta$-clustering and TLL-state both show decaying correlations.

\begin{figure}
    \centering
    \includegraphics{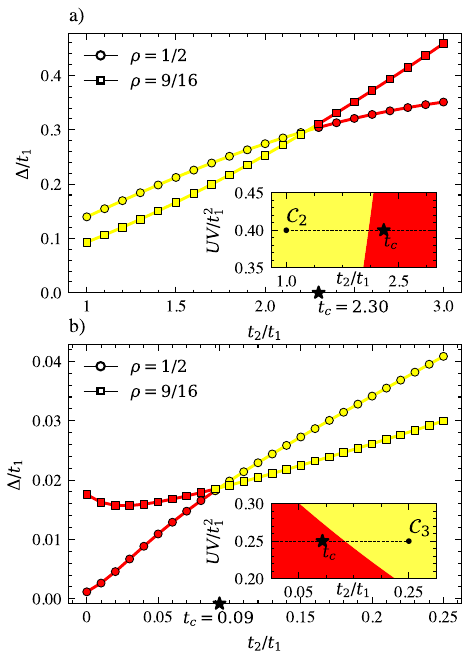}
    \caption{(a) The energy gap $\Delta/t_1$ to the half-filling ground state  
    is shown for the first excited state  
    within the same particle number sector (circles) and for the lowest energy state of the $\rho=9/16$ sector (squares) for values of $t_2/t_1$ going along $\mathcal{C}_2$ from the TLL (yellow) into the CDW phase (red) for a $L=16$ chain. Increasing $t_2/t_1$  
    leads to a level crossing with the $\rho=9/16$ state at $t_c=2.30$ marked by a black star in the perturbation theory phase diagram  
    (inset). The crossing indicates the BKT transition.  
    Since the states have different particle numbers, there is no level repulsion between them.
    (b) Same as (a) but along $\mathcal{C}_3$. The level crossing is at $t_c=0.09$. For $t_2/t_1 = 0$, the half-filling ground state is nearly degenerate $(\Delta/t_1\simeq 0)$ because the state is close to being perfectly charge density ordered.}
    \label{fig:level_spec_case2}
\end{figure}

To detect the phase transition from the CDW phase to the TLL phase, we calculated two types of 
excitation energies as a function of $t_2/t_1$ along $\mathcal{C}_{2}$ and $\mathcal{C}_{3}$ in Fig.~\ref{fig:phasediagram} and the result is shown in Fig.~\ref{fig:level_spec_case2}. One is the energy of the first excited state in the half-filling $\rho=1/2$ sector, and the other is that of the lowest energy state in the $\rho=9/16$ sector. Since the effective spin Hamiltonian Eq.~\eqref{eq:pert_xxz} has SU($2$) symmetry at the phase transition point $J_z=J_x$, these energy levels should degenerate at the transition point in the original fermionic model. By increasing $t_2/t_1$, these two energies cross at the point $t_c=2.30$ for $\mathcal{C}_{2}$ and $t_c=0.09$ for $\mathcal{C}_{3}$, which is close to the prediction of perturbation theory.
This approach, which employs the emergent SU(2) symmetry at the BKT transition point, is inspired by the method of level spectroscopy~\cite{K_Nomura_1994,ueda_resolving_2021}.

\section{Discussion}\label{sec:discussion}

In cold atom setups, both quantum many-body scarring~\cite{bernien_probing_2017,scar2, su2023observation} and SU($N$) fermions~\cite{Abraham1997,Bartenstein2005,Fukuhara2007,Ottenstein2008,Huckans2009,Taie2010,Taie2012,Desalvo2010,Lewenstein2012,Scazza2014,Zhang2014,Cazalilla2014,Pagano2014,Hofrichter2016} have been experimentally realized to great success.
However, preparing specific states higher up in energy is extremely challenging, which makes observation of ergodicity breaking  
-- in particular for fermionic systems -- elusive. As a scar state of the extended attractive SU($3$) Hubbard model, the preparation of an $\eta$-clustering state is difficult, potentially even unattainable. In particular, because it requires that the one- and two-body hopping terms are equally strong. Our extension of the attractive SU($3$) Hubbard model not only promotes the $\eta$-clustering state from an excited scar to a proper ground state, moreover it does so in a parameter range that is much more realistic for cold atom setups, i.e., small two-body hopping and nearest-neighbor attraction compared to the one-body hopping amplitude. 
We also emphasize that although we assumed that the Hamiltonian is SU($3$) symmetric for simplicity, this condition is not necessary, at least in the limit 
of large on-site attractive interaction.

Not only from a cold atom, but also from the condensed matter perspective, several theoretical avenues that extend our work present themselves. The even-$N$ molecular superfluid phases have been discussed as (strongly bound) generalizations of conventional Cooper pairs~\cite{oned1,oned2,oned3,oned4,marti_charge_2024}, however $3e$-singlets form a fermionic, not a bosonic quasi-particle. This raises the question of what type of responses a system in an $\eta$-clustering ground state might exhibit. 
While ODLRO is sufficient to enforce both the Meissner effect and flux quantization for two-particle wave functions~\cite{sewell_off-diagonal_1990,nieh_off-diagonal_1995}, it remains unexplored what impact the bODLRO has on transport. 

While superfluidity or higher-$Ne$ superconductivity are typically regarded as properties of the bulk, bODLRO -- as the name implies -- emerges at the boundaries of the system. The singlet correlation function of the $\eta$-clustering state as shown in Fig.~\ref{fig:singlet_correlation} is remarkably reminiscent of what the local density of states of a topologically non-trivial one-dimensional system looks like. In Ref.\ \cite{lang_topological_2015} a similar behavior of the correlation function in interacting fermionic ladders is found. The authors argue that the edge-edge correlations imply the existence of exponentially localized edge modes. In their setup, the ground state is gapless and degenerate due to two possible subchain-parity configurations. The $\eta$-clustering state is gapless too, and we want to emphasize that we find no direct evidence of edge modes~\footnote{If the particle number is not fixed, the ground state becomes $L$-fold degenerate due to the emergent SU(2) symmetry in the effective XXZ model with $|J_x|=|J_z|$. The $\eta$-operators $\eta$ and $\eta^\dagger$ relate each eigenmode, but they are not localized at the edges.}, which -- conventionally -- are defined for gapped systems.
Whether or not there exists a bulk invariant (other than just $N$) that allows us to infer the presence or absence of bODLRO is an open question. Alternatively, the $\eta$-clustering state might be suitably classified as a gapless symmetry-protected topological (SPT) phase~\cite{gaplessSPT}.

We leave these questions for future study and hope that our work inspires fruitful research in that direction. 

\begin{acknowledgments}
We thank Kiyohide Nomura and Masaya Nakagawa for their valuable comments. H.Y. was supported by JSPS KAKENHI Grant-in-Aid for JSPS fellows Grant No. JP22J20888, the Forefront Physics and Mathematics Program to Drive Transformation, and JSR Fellowship, the University of Tokyo.
N.H. acknowledges financial support from the Max Planck Institute for Solid State Research in Stuttgart, Germany as well as by the Japanese Society for the Promotion of Science (JSPS) and the German Academic Exchange Service (DAAD) within the JSPS Summer Program 2020 (postponed to 2022 due to COVID-19 pandemic).  
H. K. is supported by JSPS KAKENHI Grants No.\ JP23K25783, No.\ JP23K25790, and MEXT KAKENHI Grant-in-Aid for Transformative Research Areas A “Extreme Universe” (KAKENHI Grant No.\ JP21H05191).
\end{acknowledgments}

\section*{DATA AVAILABILITY}
We used open-source code bases to perform the calculations and the models used to generate all data shown in the manuscript are openly available~\cite{yoshida_edge_2024_zenodo}.

\appendix
\section{\label{sec:appendix_perturbation}
Details of Perturbation Theory}
In this section, we  
use perturbation theory for the degenerate ground states  
to derive Eqs. \eqref{eq:pert_xxz} and \eqref{eq:pert_jxjz}. 
We adopt the notation of Sec. \ref{sec:perturbation}, and write the perturbation as $\hat{H}_{\mathrm{pert}}=\hat{H}_{1}+\hat{H}_{2}+\hat{H}_{V}$. 
The first-order contribution to the effective Hamiltonian is given by $\hat{P}\hat{H}_{\mathrm{pert}}\hat{P}=\hat{P}\hat{H}_{V}\hat{P}$ because $\hat{H}_1$ and $\hat{H}_2$ have no contribution, i.e.,  
$\hat{P}\hat{H}_1\hat{P}=\hat{P}\hat{H}_2\hat{P}=0$. Since $\hat{P}\left(\hat{n}_{j}-\frac{3}{2}\right)\hat{P}=3\hat{P} \left(\hat{\eta}_{j}^{\dagger} \hat{\eta}_j-\frac{1}{2}\right) \hat{P}=3\hat{S}^z_j$, the contribution of $\hat{H}_V$ is given by 
\begin{equation}\label{eq:pert1}
\hat{P}\hat{H}_V\hat{P}=9V\sum_{j=1}^L\hat{S}^z_j\hat{S}^z_{j+1}.
\end{equation}

Next, the contribution of the second order perturbation is given by $-\hat{P} \hat{H}_{\mathrm{pert}}\left[(\hat{H}_{U}-E^\mathrm{GS}_U)^{-1} \right]^\prime\hat{H}_{\mathrm{pert}} \hat{P}$, where $E^\mathrm{GS}_U$ is the energy of the ground state of $\hat{H}_{U}$ and the $^\prime$ sign indicates that the operator $\left[(\hat{H}_{U}-E^\mathrm{GS}_U)^{-1} \right]^\prime$ acts on excited states in the natural way but vanishes on the ground states. Since $\hat{H}_{V}\mathcal{W}\subset \mathcal{W}$ and
$\left[(\hat{H}_{U}-E^\mathrm{GS}_U)^{-1} \right]^\prime \hat{H}_{i} \hat{P}=-\frac{1}{2U}\hat{H}_{i} \hat{P}$ $(i=1,2)$,
the second-order contribution is given by ${\frac{1}{2U}\hat{P} (\hat{H}_{1}+\hat{H}_{2})^2 \hat{P}}$.
Then, we find
\begin{align}\label{eq:pert2}
    &\frac{1}{2U}\hat{P} (\hat{H}_{1})^2 \hat{P} \nonumber\\
    &=\frac{t_1^2}{U}\hat{P} \sum_{j=1}^{L-1}\sum_{\alpha=1}^3 (\hat{c}^\dagger_{j+1,\alpha}\hat{c}_{j,\alpha}\hat{c}^\dagger_{j,\alpha}\hat{c}_{j+1,\alpha})\hat{P} \nonumber\\
    &=-\frac{t_1^2}{U}\hat{P} \sum_{j=1}^{L-1}\sum_{\alpha=1}^3 \left[\left(n_{j,\alpha}-\frac{1}{2}\right)\left(n_{j+1,\alpha}-\frac{1}{2}\right)-\frac{1}{4}\right]\hat{P} \nonumber\\
    &=-\frac{3t_1^2}{U}\sum_{j=1}^{L-1}\left(\hat{S}_j^z \hat{S}_{j+1}^z-\frac{1}{4}\right).
\end{align}
Similarly, we have 
\begin{align}\label{eq:pert3}
    &\frac{1}{2U}\hat{P} (\hat{H}_{2})^2 \hat{P}=-\frac{3t_2^2}{U}\sum_{j=1}^{L-1}\left(\hat{S}_j^z \hat{S}_{j+1}^z-\frac{1}{4}\right).
\end{align}
Finally, 
\begin{align}\label{eq:pert4}
    &\frac{1}{2U}\hat{P} \hat{H}_{1} \hat{H}_{2} \hat{P}=\frac{1}{2U}\hat{P} \hat{H}_{2} \hat{H}_{1} \hat{P} \nonumber\\
    &=-\frac{t_1 t_2}{2U}\hat{P} \sum_{j=1}^{L-1}\sum_{\alpha=1}^3 (\hat{c}^\dagger_{j+1,\alpha}\hat{c}_{j,\alpha}\hat{\bar{c}}^\dagger_{j+1,\alpha}\hat{\bar{c}}_{j,\alpha}+\text{h.c.})\hat{P} \nonumber\\
    &=-\frac{3t_1 t_2}{2U}\hat{P} \sum_{j=1}^{L-1}(\hat{\eta}^\dagger_{j+1}\hat{\eta}_{j}+\hat{\eta}^\dagger_{j}\hat{\eta}_{j+1})\hat{P} \nonumber\\
    &=\frac{3t_1 t_2}{2U}\sum_{j=1}^{L-1} \left(\hat{S}_j^+ \hat{S}_{j+1}^-+\hat{S}_j^- \hat{S}_{j+1}^+ \right).
\end{align}
By summarizing Eqs. \eqref{eq:pert1}, \eqref{eq:pert2}, \eqref{eq:pert3}, and \eqref{eq:pert4}, we obtain Eq.~\eqref{eq:pert_xxz} up to 
a constant energy shift.

\section{\label{sec:fragmentation} Hilbert-space fragmentation at $t_1=0$}
The model~\eqref{eq:hamiltonian} with $t_1=0$ exhibits Hilbert-space fragmentation~\cite{sala_ergodicity_2020,khemani_localization_2020,zadnik_folded_2021,pozsgay2021integrable,moudgalya_thermalization_2021, scar1}. Furthermore, in the subspace spanned by states without singly or triply occupied sites, which we denote by $\mathcal{X}$, each fragmented sector can be exactly described by the spin-$1/2$ XXZ Hamiltonian. In this subspace, we write $\ket{0}_j$ the empty state at site $j$, and $\ket{\bar{\alpha}}_j=\hat{\bar{c}}^\dagger_{j,\alpha}\ket{0}_j$ for $\alpha=1,2,3$ the doubly occupied states. Then, the Hamiltonian~\eqref{eq:hamiltonian} projected down to $\mathcal{X}$ is
\begin{align}
\begin{split}
    \hat{H}_\mathrm{eff}=&-t_2 \sum_{j=1}^{L-1} \sum_{\alpha=1}^3 \left( \ket{\bar{\alpha},0}_{j,j+1}\bra{0,\bar{\alpha}}+\mathrm{h.c.}\right)\\
    &+V \sum_{j=1}^{L-1}\sum_{\alpha=1}^3 \ket{\bar{\alpha},\bar{\beta}}_{j,j+1}\bra{\bar{\alpha},\bar{\beta}}\\
    &+3V \sum_{j=1}^{L-1} \ket{0,0}_{j,j+1}\bra{0,0}+\text{const.},
    \end{split}
\end{align}
where the constant depends on the total fermion number and the number of doubly occupied sites. This model turns out to be a special case of Maassarani's XXC model~\cite{maassarani_xxc_1998, arnaudon1998integrable}. This model is integrable and exhibits Hilbert-space fragmentation, which is easy to see by noting that swaps of $\bar{\alpha}$ and $\bar{\beta}$ never happen with $\hat{H}_\mathrm{eff}$. In other words, the arrangement of $\bar{\alpha}$'s obtained by disregarding the $0$'s in $\ket{0,\ldots,0,\bar{\alpha}_1,0,\ldots,0,\bar{\alpha}_2,0,\ldots}$ is conserved.

With this in mind, we pick one of the arrangements of $\bar{\alpha}$'s and fix it. Then, the effective Hamiltonian within this sector is 
\begin{align}   
\begin{split}
\hat{\tilde{H}}_\mathrm{eff}=&-t_2 \sum_{j=1}^{L-1} \left( \ket{1,0}_{j,j+1}\bra{0,1}+\mathrm{h.c.}\right)\\
&+V \sum_{j=1}^{L-1} \ket{1,1}_{j,j+1}\bra{1,1}\\
&+3V \sum_{j=1}^{L-1} \ket{0,0}_{j,j+1}\bra{0,0}+\text{const.},
\end{split}
\label{eq:effective_2}
\end{align}
where we disregard the difference of $\ket{\bar{1}}$, $\ket{\bar{2}}$, and $\ket{\bar{3}}$ and simply identify them as $\ket{1}$. 
Introducing effective spin operators $\hat{\sigma}^x=\ket{1} \bra{0}+\ket{0} \bra{1}$, $\hat{\sigma}^y=-i(\ket{1}\bra{0}-\ket{0}\bra{1})$, and $\hat{\sigma}^z_j=\ket{1}\bra{1}-\ket{0} \bra{0}$, Eq. \eqref{eq:effective_2} can be written as
\begin{align}   
\begin{split}
\hat{\tilde{H}}_\mathrm{eff}=\sum_{j=1}^{L-1}& \bigg[-2t_2 \left(\hat{\sigma}_j^x\hat{\sigma}_{j+1}^x + \hat{\sigma}_j^y\hat{\sigma}_{j+1}^y\right)+V\hat{\sigma}_j^z\hat{\sigma}_{j+1}^z \\
&-\frac{V}{2}(\hat{\sigma}_j^z+\hat{\sigma}_{j+1}^z) 
\bigg]+\text{const.},
\end{split}
\label{eq:effective_3}
\end{align}
which is the XXZ model in a magnetic field.

\begin{figure}
    \centering
    \includegraphics[width=\columnwidth]{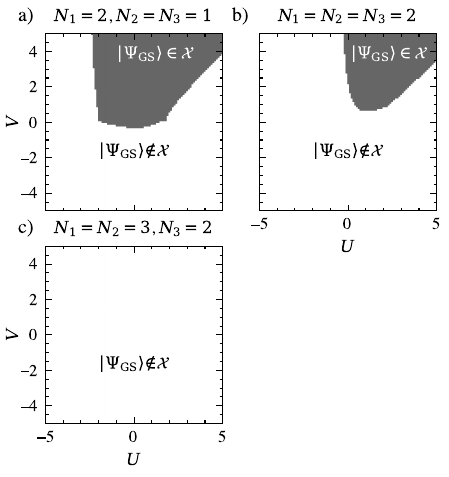}
    \caption{Numerical results showing whether the ground state belongs to the subspace $\mathcal{X}$.
    The black (white) region indicates $\ket{\Psi_{\text{GS}}}\in \mathcal{X}$ ($\ket{\Psi_{\text{GS}}}\notin \mathcal{X}$) for system size $L=8$ and particle numbers (a) $N_1=2$, $N_2=N_3=1$, (b) $N_1=N_2=N_3=2$ and (c) $N_1=N_2=3$, $N_3=2$.}
    \label{fig:fragmented}
\end{figure}

We also numerically investigated the parameter conditions  
under which the ground state $\ket{\Psi_{\text{GS}}}$ belongs to the subspace $\mathcal{X}$. The results for a system size $L=8$ and $t_2=1$ are shown in Fig.~\ref{fig:fragmented}. The number of particles is fixed for each flavor, and they are denoted as $N_\alpha$ $(\alpha=1,2,3)$. 
Numerical results suggest that in a certain parameter region, the ground state lies in the subspace $\mathcal{X}$, in which case the ground-state properties can be analyzed analytically by exploiting the partial integrability of the model.

\section{\label{sec:appendix_numerical_methods} Numerical Methods}

\subsection{\label{app:dmrg}DMRG calculations}

The Density Matrix Renormalization Group (DMRG) algorithm is a powerful variational numerical method for ground state searches of one-dimensional and quasi-one-dimensional systems \cite{white1992density}. The calculations presented in this work have been performed as implemented in the TeNPy package \cite{hauschild2018efficient}.

An SU($3$) site can be represented by 
\begin{align}
|s_j\rangle = \bigotimes_{\alpha=1}^{3}|s_j^\alpha\rangle
\end{align}
where $|s_j\rangle$ is a spinless fermion site with local Hilbert space $|s_j^\alpha\rangle = \{ |\tikzcircle[black, fill=white]{3pt}\rangle, |\tikzcircle[black, fill=black]{3pt}\rangle \} $
and flavor $\alpha = {1, 2 ,3}$. 
The filled and open circles represent occupied and unoccupied sites, respectively.
In this representation, the longest interaction is between nearest neighbors and the local dimension of the Hilbert space is $8$. Because DMRG scales cubic in the dimension of the local Hilbert space, but linear in system size, we represent an SU($3$) chain of $L$ sites as a chain of $3L$ spinless fermionic sites. This comes, however, at the cost of introducing up to fifth-nearest neighbor interaction, which requires larger bond dimension of the Matrix Product Operator (MPO), and makes the evaluation of correlation functions in general more costly~\cite{schollwock2011density}.

In the basis of $3L$ spinless fermions, we choose an initial product state in the half-filling sector given by

\begin{align}\label{eq:prod_state_init}
    |\Psi_{\rm init}^\otimes\rangle = \bigotimes_{j=1}^{L/2}|\tikzcircle[black, fill=white]{3pt} \hspace{0.6pt}\tikzcircle[black, fill=white]{3pt} \hspace{0.6pt}\tikzcircle[black, fill=white]{3pt}\rangle_{2j-1}  \otimes | \tikzcircle[black, fill=black]{3pt} \hspace{0.6pt}\tikzcircle[black, fill=black]{3pt} \hspace{0.6pt}\tikzcircle[black, fill=black]{3pt}\rangle_{2j}.
\end{align}
with an even number of SU($3$) sites $L$.

This state is a charge density wave with $\pi$-ordering that is two-fold degenerate. To build a less biased initial state, we construct random unitaries $\hat{U}^{\rm rand}$ that we use to `time-evolve' the state:
\begin{align}
    |\Psi_{\rm init}\rangle = \prod^{10}_{i=1}\hat{U}_i^{\rm rand}(\Delta t = 1/t_1)|\Psi_{\rm init}^\otimes\rangle 
\end{align}

We now have a Matrix Product State (MPS) of the general form 
\begin{align}
    |\Psi\rangle = \sum_{\substack{s_1, \ldots,s_{3L}\\ \alpha_2, \ldots,\alpha_{3L}}}M^{[1]s_1}_{\alpha_1\alpha_2}M^{[2]s_2}_{\alpha_2\alpha_3}\ldots M^{[3L]s_{3L}}_{\alpha_{3L}\alpha_{3L+1}}|s_1, s_2,\ldots, s_{3L}\rangle,
\end{align}
where each $M^{\left[j \right]s_{j}}$ is a $\chi_j\times \chi_{j+1}$ matrix and $L$ the number of SU(3) sites~\cite{hauschild2018efficient}, where $\chi$ is the bond dimension. We use the common two-site update scheme, which sweeps through the system and iteratively optimizes the matrices by minimizing the energy locally on two sites. The procedure is repeated until the convergence criteria are fulfilled:
\begin{align}
\Delta E &< 10^{-8}t_1 \\  \Delta S_A &< 10^{-6}.    
\end{align}
In terms of MPS the entanglement entropy $S_A$ for a bipartition $A$ is computed as
\begin{align}
    S_A = -\sum_j \Lambda_j ^2\log \Lambda_j^2
\end{align}
where $\Lambda$ is a diagonal matrix that contains the Schmidt values~\cite{hauschild2018efficient}.

To obtain the first excited state of the half-filling sector $|\Psi_1^{\rho=1/2}\rangle$, we first obtain the ground state of that sector $|\Psi_0^{\rho=1/2}\rangle$. Then DMRG is run again, this time with the additional orthogonality constraint
\begin{align}
    \langle \Psi_1^{\rho=1/2}|\Psi_0^{\rho=1/2}\rangle = 0.
\end{align}
To find the excitations away from half-filling, we simply add or subtract the desired number of occupied SU($3$)-sites to our initial product state from \mbox{Eq.\ \eqref{eq:prod_state_init}}, for example: 
\begin{align}
    &\hat{\eta}^\dagger_{1} |\Psi_{\rm init}^{\otimes,\rho=1/2}\rangle \\&=|\tikzcircle[black, fill=black]{3pt} \hspace{0.6pt}\tikzcircle[black, fill=black]{3pt} \hspace{0.6pt}\tikzcircle[black, fill=black]{3pt}\rangle_1 \otimes \left[\bigotimes_{j=1}^{L/2 - 1} | \tikzcircle[black, fill=black]{3pt} \hspace{0.6pt}\tikzcircle[black, fill=black]{3pt} \hspace{0.6pt}\tikzcircle[black, fill=black]{3pt}\rangle_{2j} \otimes |\tikzcircle[black, fill=white]{3pt} \hspace{0.6pt}\tikzcircle[black, fill=white]{3pt} \hspace{0.6pt}\tikzcircle[black, fill=white]{3pt}\rangle_{2j+1} \right] \otimes |\tikzcircle[black, fill=black]{3pt} \hspace{0.6pt}\tikzcircle[black, fill=black]{3pt} \hspace{0.6pt}\tikzcircle[black, fill=black]{3pt}\rangle_L
\end{align}
Because the Hamiltonian \mbox{(see Eq.\ \eqref{eq:hamiltonian})} preserves particle number, we know that the resulting ground state must lie within the same filling sector as the initial state.

\bibliography{references}

\begin{thebibliography}{109}%
\makeatletter
\providecommand \@ifxundefined [1]{%
 \@ifx{#1\undefined}
}%
\providecommand \@ifnum [1]{%
 \ifnum #1\expandafter \@firstoftwo
 \else \expandafter \@secondoftwo
 \fi
}%
\providecommand \@ifx [1]{%
 \ifx #1\expandafter \@firstoftwo
 \else \expandafter \@secondoftwo
 \fi
}%
\providecommand \natexlab [1]{#1}%
\providecommand \enquote  [1]{``#1''}%
\providecommand \bibnamefont  [1]{#1}%
\providecommand \bibfnamefont [1]{#1}%
\providecommand \citenamefont [1]{#1}%
\providecommand \href@noop [0]{\@secondoftwo}%
\providecommand \href [0]{\begingroup \@sanitize@url \@href}%
\providecommand \@href[1]{\@@startlink{#1}\@@href}%
\providecommand \@@href[1]{\endgroup#1\@@endlink}%
\providecommand \@sanitize@url [0]{\catcode `\\12\catcode `\$12\catcode `\&12\catcode `\#12\catcode `\^12\catcode `\_12\catcode `\%12\relax}%
\providecommand \@@startlink[1]{}%
\providecommand \@@endlink[0]{}%
\providecommand \url  [0]{\begingroup\@sanitize@url \@url }%
\providecommand \@url [1]{\endgroup\@href {#1}{\urlprefix }}%
\providecommand \urlprefix  [0]{URL }%
\providecommand \Eprint [0]{\href }%
\providecommand \doibase [0]{https://doi.org/}%
\providecommand \selectlanguage [0]{\@gobble}%
\providecommand \bibinfo  [0]{\@secondoftwo}%
\providecommand \bibfield  [0]{\@secondoftwo}%
\providecommand \translation [1]{[#1]}%
\providecommand \BibitemOpen [0]{}%
\providecommand \bibitemStop [0]{}%
\providecommand \bibitemNoStop [0]{.\EOS\space}%
\providecommand \EOS [0]{\spacefactor3000\relax}%
\providecommand \BibitemShut  [1]{\csname bibitem#1\endcsname}%
\let\auto@bib@innerbib\@empty
\bibitem [{\citenamefont {Yang}(1989)}]{yang_1989_eta}%
  \BibitemOpen
  \bibfield  {author} {\bibinfo {author} {\bibfnamefont {C.~N.}\ \bibnamefont {Yang}},\ }\bibfield  {title} {\bibinfo {title} {$\eta$ pairing and off-diagonal long-range order in a {Hubbard} model},\ }\href {https://doi.org/10.1103/PhysRevLett.63.2144} {\bibfield  {journal} {\bibinfo  {journal} {Phys. Rev. Lett.}\ }\textbf {\bibinfo {volume} {63}},\ \bibinfo {pages} {2144} (\bibinfo {year} {1989})}\BibitemShut {NoStop}%
\bibitem [{\citenamefont {Yang}(1962)}]{yang_concept_1962}%
  \BibitemOpen
  \bibfield  {author} {\bibinfo {author} {\bibfnamefont {C.~N.}\ \bibnamefont {Yang}},\ }\bibfield  {title} {\bibinfo {title} {Concept of {Off}-{Diagonal} {Long}-{Range} {Order} and the {Quantum} {Phases} of {Liquid} {He} and of {Superconductors}},\ }\href {https://doi.org/10.1103/RevModPhys.34.694} {\bibfield  {journal} {\bibinfo  {journal} {Rev. Mod. Phys.}\ }\textbf {\bibinfo {volume} {34}},\ \bibinfo {pages} {694} (\bibinfo {year} {1962})}\BibitemShut {NoStop}%
\bibitem [{\citenamefont {Sewell}(1990)}]{sewell_off-diagonal_1990}%
  \BibitemOpen
  \bibfield  {author} {\bibinfo {author} {\bibfnamefont {G.~L.}\ \bibnamefont {Sewell}},\ }\bibfield  {title} {\bibinfo {title} {Off-diagonal long-range order and the {Meissner} effect},\ }\href {https://doi.org/10.1007/BF01013973} {\bibfield  {journal} {\bibinfo  {journal} {J. Stat. Phys.}\ }\textbf {\bibinfo {volume} {61}},\ \bibinfo {pages} {415} (\bibinfo {year} {1990})}\BibitemShut {NoStop}%
\bibitem [{\citenamefont {Nieh}\ \emph {et~al.}(1995)\citenamefont {Nieh}, \citenamefont {Su},\ and\ \citenamefont {Zhao}}]{nieh_off-diagonal_1995}%
  \BibitemOpen
  \bibfield  {author} {\bibinfo {author} {\bibfnamefont {H.~T.}\ \bibnamefont {Nieh}}, \bibinfo {author} {\bibfnamefont {G.}~\bibnamefont {Su}},\ and\ \bibinfo {author} {\bibfnamefont {B.-H.}\ \bibnamefont {Zhao}},\ }\bibfield  {title} {\bibinfo {title} {Off-diagonal long-range order: {Meissner} effect and flux quantization},\ }\href {https://doi.org/10.1103/PhysRevB.51.3760} {\bibfield  {journal} {\bibinfo  {journal} {Phys. Rev. B}\ }\textbf {\bibinfo {volume} {51}},\ \bibinfo {pages} {3760} (\bibinfo {year} {1995})}\BibitemShut {NoStop}%
\bibitem [{\citenamefont {Kitamura}\ and\ \citenamefont {Aoki}(2016)}]{kitamura_eta_2016}%
  \BibitemOpen
  \bibfield  {author} {\bibinfo {author} {\bibfnamefont {S.}~\bibnamefont {Kitamura}}\ and\ \bibinfo {author} {\bibfnamefont {H.}~\bibnamefont {Aoki}},\ }\bibfield  {title} {\bibinfo {title} {$\eta$-pairing superfluid in periodically-driven fermionic {Hubbard} model with strong attraction},\ }\href {https://doi.org/10.1103/PhysRevB.94.174503} {\bibfield  {journal} {\bibinfo  {journal} {Phys. Rev. B}\ }\textbf {\bibinfo {volume} {94}},\ \bibinfo {pages} {174503} (\bibinfo {year} {2016})}\BibitemShut {NoStop}%
\bibitem [{\citenamefont {Kaneko}\ \emph {et~al.}(2019)\citenamefont {Kaneko}, \citenamefont {Shirakawa}, \citenamefont {Sorella},\ and\ \citenamefont {Yunoki}}]{kaneko_photoinduced_2019}%
  \BibitemOpen
  \bibfield  {author} {\bibinfo {author} {\bibfnamefont {T.}~\bibnamefont {Kaneko}}, \bibinfo {author} {\bibfnamefont {T.}~\bibnamefont {Shirakawa}}, \bibinfo {author} {\bibfnamefont {S.}~\bibnamefont {Sorella}},\ and\ \bibinfo {author} {\bibfnamefont {S.}~\bibnamefont {Yunoki}},\ }\bibfield  {title} {\bibinfo {title} {Photoinduced $\eta$ {Pairing} in the {Hubbard} {Model}},\ }\href {https://doi.org/10.1103/PhysRevLett.122.077002} {\bibfield  {journal} {\bibinfo  {journal} {Phys. Rev. Lett.}\ }\textbf {\bibinfo {volume} {122}},\ \bibinfo {pages} {077002} (\bibinfo {year} {2019})}\BibitemShut {NoStop}%
\bibitem [{\citenamefont {Bu{\v{c}}a}\ \emph {et~al.}(2019)\citenamefont {Bu{\v{c}}a}, \citenamefont {Tindall},\ and\ \citenamefont {Jaksch}}]{buvca_eta_2019}%
  \BibitemOpen
  \bibfield  {author} {\bibinfo {author} {\bibfnamefont {B.}~\bibnamefont {Bu{\v{c}}a}}, \bibinfo {author} {\bibfnamefont {J.}~\bibnamefont {Tindall}},\ and\ \bibinfo {author} {\bibfnamefont {D.}~\bibnamefont {Jaksch}},\ }\bibfield  {title} {\bibinfo {title} {Non-stationary coherent quantum many-body dynamics through dissipation},\ }\href@noop {} {\bibfield  {journal} {\bibinfo  {journal} {Nat. Commun.}\ }\textbf {\bibinfo {volume} {10}},\ \bibinfo {pages} {1} (\bibinfo {year} {2019})}\BibitemShut {NoStop}%
\bibitem [{\citenamefont {Tindall}\ \emph {et~al.}(2019)\citenamefont {Tindall}, \citenamefont {Bu\ifmmode~\check{c}\else \v{c}\fi{}a}, \citenamefont {Coulthard},\ and\ \citenamefont {Jaksch}}]{tindall_eta_2019}%
  \BibitemOpen
  \bibfield  {author} {\bibinfo {author} {\bibfnamefont {J.}~\bibnamefont {Tindall}}, \bibinfo {author} {\bibfnamefont {B.}~\bibnamefont {Bu\ifmmode~\check{c}\else \v{c}\fi{}a}}, \bibinfo {author} {\bibfnamefont {J.~R.}\ \bibnamefont {Coulthard}},\ and\ \bibinfo {author} {\bibfnamefont {D.}~\bibnamefont {Jaksch}},\ }\bibfield  {title} {\bibinfo {title} {{Heating-Induced Long-Range $\eta$ Pairing in the Hubbard Model}},\ }\href {https://doi.org/10.1103/PhysRevLett.123.030603} {\bibfield  {journal} {\bibinfo  {journal} {Phys. Rev. Lett.}\ }\textbf {\bibinfo {volume} {123}},\ \bibinfo {pages} {030603} (\bibinfo {year} {2019})}\BibitemShut {NoStop}%
\bibitem [{\citenamefont {Li}(2020)}]{li_eta-pairing_2020}%
  \BibitemOpen
  \bibfield  {author} {\bibinfo {author} {\bibfnamefont {K.}~\bibnamefont {Li}},\ }\bibfield  {title} {\bibinfo {title} {$\eta$-pairing in correlated fermion models with spin-orbit coupling},\ }\href {https://doi.org/10.1103/PhysRevB.102.165150} {\bibfield  {journal} {\bibinfo  {journal} {Phys. Rev. B}\ }\textbf {\bibinfo {volume} {102}},\ \bibinfo {pages} {165150} (\bibinfo {year} {2020})}\BibitemShut {NoStop}%
\bibitem [{\citenamefont {Tsuji}\ \emph {et~al.}(2021)\citenamefont {Tsuji}, \citenamefont {Nakagawa},\ and\ \citenamefont {Ueda}}]{tsuji_tachyonic_2021}%
  \BibitemOpen
  \bibfield  {author} {\bibinfo {author} {\bibfnamefont {N.}~\bibnamefont {Tsuji}}, \bibinfo {author} {\bibfnamefont {M.}~\bibnamefont {Nakagawa}},\ and\ \bibinfo {author} {\bibfnamefont {M.}~\bibnamefont {Ueda}},\ }\bibfield  {title} {\bibinfo {title} {Tachyonic and {Plasma} {Instabilities} of $\eta$-{Pairing} {States} {Coupled} to {Electromagnetic} {Fields}},\ }\href {http://arxiv.org/abs/2103.01547} {\bibfield  {journal} {\bibinfo  {journal} {arXiv: 2103.01547}\ } (\bibinfo {year} {2021})}\BibitemShut {NoStop}%
\bibitem [{\citenamefont {Nakagawa}\ \emph {et~al.}(2021)\citenamefont {Nakagawa}, \citenamefont {Tsuji}, \citenamefont {Kawakami},\ and\ \citenamefont {Ueda}}]{nakagawa_eta_2021}%
  \BibitemOpen
  \bibfield  {author} {\bibinfo {author} {\bibfnamefont {M.}~\bibnamefont {Nakagawa}}, \bibinfo {author} {\bibfnamefont {N.}~\bibnamefont {Tsuji}}, \bibinfo {author} {\bibfnamefont {N.}~\bibnamefont {Kawakami}},\ and\ \bibinfo {author} {\bibfnamefont {M.}~\bibnamefont {Ueda}},\ }\bibfield  {title} {\bibinfo {title} {$\eta$ {Pairing} of {Light}-{Emitting} {Fermions}: {Nonequilibrium} {Pairing} {Mechanism} at {High} {Temperatures}},\ }\href {http://arxiv.org/abs/2103.13624} {\bibfield  {journal} {\bibinfo  {journal} {arXiv: 2103.13624}\ } (\bibinfo {year} {2021})}\BibitemShut {NoStop}%
\bibitem [{\citenamefont {Moudgalya}\ \emph {et~al.}(2022)\citenamefont {Moudgalya}, \citenamefont {Bernevig},\ and\ \citenamefont {Regnault}}]{scar1}%
  \BibitemOpen
  \bibfield  {author} {\bibinfo {author} {\bibfnamefont {S.}~\bibnamefont {Moudgalya}}, \bibinfo {author} {\bibfnamefont {B.~A.}\ \bibnamefont {Bernevig}},\ and\ \bibinfo {author} {\bibfnamefont {N.}~\bibnamefont {Regnault}},\ }\bibfield  {title} {\bibinfo {title} {{Quantum many-body scars and Hilbert space fragmentation: a review of exact results}},\ }\href {https://iopscience.iop.org/article/10.1088/1361-6633/ac73a0/meta?casa_token=-Mk3T6bogCcAAAAA:0pSifvQI5UqMtmtREl4LF7ieofByjZ3lIok2rQKQUS74UDBr3y5UUYnVT7Is9fKRFNW9gTUhcKdMQesaPXviFvyf3Wcw} {\bibfield  {journal} {\bibinfo  {journal} {Rep. Prog. Phys.}\ }\textbf {\bibinfo {volume} {85}},\ \bibinfo {pages} {086501} (\bibinfo {year} {2022})}\BibitemShut {NoStop}%
\bibitem [{\citenamefont {Serbyn}\ \emph {et~al.}(2021)\citenamefont {Serbyn}, \citenamefont {Abanin},\ and\ \citenamefont {Papi{\'{c}}}}]{scar2}%
  \BibitemOpen
  \bibfield  {author} {\bibinfo {author} {\bibfnamefont {M.}~\bibnamefont {Serbyn}}, \bibinfo {author} {\bibfnamefont {D.~A.}\ \bibnamefont {Abanin}},\ and\ \bibinfo {author} {\bibfnamefont {Z.}~\bibnamefont {Papi{\'{c}}}},\ }\bibfield  {title} {\bibinfo {title} {Quantum many-body scars and weak breaking of ergodicity},\ }\href {https://doi.org/10.1038/s41567-021-01230-2} {\bibfield  {journal} {\bibinfo  {journal} {Nat. Phys.}\ }\textbf {\bibinfo {volume} {17}},\ \bibinfo {pages} {675} (\bibinfo {year} {2021})}\BibitemShut {NoStop}%
\bibitem [{\citenamefont {Turner}\ \emph {et~al.}(2018)\citenamefont {Turner}, \citenamefont {Michailidis}, \citenamefont {Abanin}, \citenamefont {Serbyn},\ and\ \citenamefont {Papi{\'{c}}}}]{scar3}%
  \BibitemOpen
  \bibfield  {author} {\bibinfo {author} {\bibfnamefont {C.~J.}\ \bibnamefont {Turner}}, \bibinfo {author} {\bibfnamefont {A.~A.}\ \bibnamefont {Michailidis}}, \bibinfo {author} {\bibfnamefont {D.~A.}\ \bibnamefont {Abanin}}, \bibinfo {author} {\bibfnamefont {M.}~\bibnamefont {Serbyn}},\ and\ \bibinfo {author} {\bibfnamefont {Z.}~\bibnamefont {Papi{\'{c}}}},\ }\bibfield  {title} {\bibinfo {title} {Weak ergodicity breaking from quantum many-body scars},\ }\href {https://doi.org/10.1038/s41567-018-0137-5} {\bibfield  {journal} {\bibinfo  {journal} {Nat. Phys.}\ }\textbf {\bibinfo {volume} {14}},\ \bibinfo {pages} {745} (\bibinfo {year} {2018})}\BibitemShut {NoStop}%
\bibitem [{\citenamefont {Chandran}\ \emph {et~al.}(2023)\citenamefont {Chandran}, \citenamefont {Iadecola}, \citenamefont {Khemani},\ and\ \citenamefont {Moessner}}]{scar4}%
  \BibitemOpen
  \bibfield  {author} {\bibinfo {author} {\bibfnamefont {A.}~\bibnamefont {Chandran}}, \bibinfo {author} {\bibfnamefont {T.}~\bibnamefont {Iadecola}}, \bibinfo {author} {\bibfnamefont {V.}~\bibnamefont {Khemani}},\ and\ \bibinfo {author} {\bibfnamefont {R.}~\bibnamefont {Moessner}},\ }\bibfield  {title} {\bibinfo {title} {Quantum many-body scars: A quasiparticle perspective},\ }\href {https://doi.org/https://doi.org/10.1146/annurev-conmatphys-031620-101617} {\bibfield  {journal} {\bibinfo  {journal} {Annu. Rev. Condens. Matter Phys.}\ }\textbf {\bibinfo {volume} {14}},\ \bibinfo {pages} {443} (\bibinfo {year} {2023})}\BibitemShut {NoStop}%
\bibitem [{\citenamefont {Mark}\ and\ \citenamefont {Motrunich}(2020)}]{mark_eta_2020}%
  \BibitemOpen
  \bibfield  {author} {\bibinfo {author} {\bibfnamefont {D.~K.}\ \bibnamefont {Mark}}\ and\ \bibinfo {author} {\bibfnamefont {O.~I.}\ \bibnamefont {Motrunich}},\ }\bibfield  {title} {\bibinfo {title} {{$\ensuremath{\eta}$-pairing states as true scars in an extended Hubbard model}},\ }\href {https://doi.org/10.1103/PhysRevB.102.075132} {\bibfield  {journal} {\bibinfo  {journal} {Phys. Rev. B}\ }\textbf {\bibinfo {volume} {102}},\ \bibinfo {pages} {075132} (\bibinfo {year} {2020})}\BibitemShut {NoStop}%
\bibitem [{\citenamefont {Moudgalya}\ \emph {et~al.}(2020)\citenamefont {Moudgalya}, \citenamefont {Regnault},\ and\ \citenamefont {Bernevig}}]{moudgalya_2020_eta}%
  \BibitemOpen
  \bibfield  {author} {\bibinfo {author} {\bibfnamefont {S.}~\bibnamefont {Moudgalya}}, \bibinfo {author} {\bibfnamefont {N.}~\bibnamefont {Regnault}},\ and\ \bibinfo {author} {\bibfnamefont {B.~A.}\ \bibnamefont {Bernevig}},\ }\bibfield  {title} {\bibinfo {title} {$\eta$-pairing in {Hubbard} models: {From} spectrum generating algebras to quantum many-body scars},\ }\href {https://doi.org/10.1103/PhysRevB.102.085140} {\bibfield  {journal} {\bibinfo  {journal} {Phys. Rev. B}\ }\textbf {\bibinfo {volume} {102}},\ \bibinfo {pages} {085140} (\bibinfo {year} {2020})}\BibitemShut {NoStop}%
\bibitem [{\citenamefont {Pakrouski}\ \emph {et~al.}(2020)\citenamefont {Pakrouski}, \citenamefont {Pallegar}, \citenamefont {Popov},\ and\ \citenamefont {Klebanov}}]{pakrouski_many-body_2020}%
  \BibitemOpen
  \bibfield  {author} {\bibinfo {author} {\bibfnamefont {K.}~\bibnamefont {Pakrouski}}, \bibinfo {author} {\bibfnamefont {P.~N.}\ \bibnamefont {Pallegar}}, \bibinfo {author} {\bibfnamefont {F.~K.}\ \bibnamefont {Popov}},\ and\ \bibinfo {author} {\bibfnamefont {I.~R.}\ \bibnamefont {Klebanov}},\ }\bibfield  {title} {\bibinfo {title} {Many-{Body} {Scars} as a {Group} {Invariant} {Sector} of {Hilbert} {Space}},\ }\href {https://doi.org/10.1103/PhysRevLett.125.230602} {\bibfield  {journal} {\bibinfo  {journal} {Phys. Rev. Lett.}\ }\textbf {\bibinfo {volume} {125}},\ \bibinfo {pages} {230602} (\bibinfo {year} {2020})}\BibitemShut {NoStop}%
\bibitem [{\citenamefont {Pakrouski}\ \emph {et~al.}(2021)\citenamefont {Pakrouski}, \citenamefont {Pallegar}, \citenamefont {Popov},\ and\ \citenamefont {Klebanov}}]{pakrouski_group_2021}%
  \BibitemOpen
  \bibfield  {author} {\bibinfo {author} {\bibfnamefont {K.}~\bibnamefont {Pakrouski}}, \bibinfo {author} {\bibfnamefont {P.~N.}\ \bibnamefont {Pallegar}}, \bibinfo {author} {\bibfnamefont {F.~K.}\ \bibnamefont {Popov}},\ and\ \bibinfo {author} {\bibfnamefont {I.~R.}\ \bibnamefont {Klebanov}},\ }\bibfield  {title} {\bibinfo {title} {Group theoretic approach to many-body scar states in fermionic lattice models},\ }\href {http://arxiv.org/abs/2106.10300} {\bibfield  {journal} {\bibinfo  {journal} {arXiv: 2106.10300}\ } (\bibinfo {year} {2021})}\BibitemShut {NoStop}%
\bibitem [{\citenamefont {Zhai}(2005)}]{zhai2005two}%
  \BibitemOpen
  \bibfield  {author} {\bibinfo {author} {\bibfnamefont {H.}~\bibnamefont {Zhai}},\ }\bibfield  {title} {\bibinfo {title} {{Two generalizations of $\eta$ pairing in extended Hubbard models}},\ }\href {https://doi.org/10.1103/PhysRevB.71.012512} {\bibfield  {journal} {\bibinfo  {journal} {Phys. Rev. B}\ }\textbf {\bibinfo {volume} {71}},\ \bibinfo {pages} {012512} (\bibinfo {year} {2005})}\BibitemShut {NoStop}%
\bibitem [{\citenamefont {Yoshida}\ and\ \citenamefont {Katsura}(2022)}]{yoshida_exact_2022}%
  \BibitemOpen
  \bibfield  {author} {\bibinfo {author} {\bibfnamefont {H.}~\bibnamefont {Yoshida}}\ and\ \bibinfo {author} {\bibfnamefont {H.}~\bibnamefont {Katsura}},\ }\bibfield  {title} {\bibinfo {title} {Exact eigenstates of extended {SU($N$)} {Hubbard} models: Generalization of {$\eta$}-pairing states with {$N$}-particle off-diagonal long-range order},\ }\href {https://doi.org/10.1103/PhysRevB.105.024520} {\bibfield  {journal} {\bibinfo  {journal} {Phys. Rev. B}\ }\textbf {\bibinfo {volume} {105}},\ \bibinfo {pages} {024520} (\bibinfo {year} {2022})}\BibitemShut {NoStop}%
\bibitem [{\citenamefont {Nakagawa}\ \emph {et~al.}(2022)\citenamefont {Nakagawa}, \citenamefont {Katsura},\ and\ \citenamefont {Ueda}}]{nakagawa_exact_2022}%
  \BibitemOpen
  \bibfield  {author} {\bibinfo {author} {\bibfnamefont {M.}~\bibnamefont {Nakagawa}}, \bibinfo {author} {\bibfnamefont {H.}~\bibnamefont {Katsura}},\ and\ \bibinfo {author} {\bibfnamefont {M.}~\bibnamefont {Ueda}},\ }\bibfield  {title} {\bibinfo {title} {Exact eigenstates of multicomponent {Hubbard} models: {SU($N$)} magnetic $\eta$ pairing, weak ergodicity breaking, and partial integrability},\ }\href {http://arxiv.org/abs/2205.07235} {\bibfield  {journal} {\bibinfo  {journal} {arXiv:2205.07235}\ } (\bibinfo {year} {2022})}\BibitemShut {NoStop}%
\bibitem [{\citenamefont {Imai}\ and\ \citenamefont {Tsuji}(2024)}]{imai_2024_systematic}%
  \BibitemOpen
  \bibfield  {author} {\bibinfo {author} {\bibfnamefont {S.}~\bibnamefont {Imai}}\ and\ \bibinfo {author} {\bibfnamefont {N.}~\bibnamefont {Tsuji}},\ }\bibfield  {title} {\bibinfo {title} {A systematic framework to construct unconventional superconducting pairing scar states using multi-body interactions},\ }\href@noop {} {\bibfield  {journal} {\bibinfo  {journal} {arXiv:2404.02914}\ } (\bibinfo {year} {2024})}\BibitemShut {NoStop}%
\bibitem [{\citenamefont {Kitaev}(2001)}]{1dinterac2}%
  \BibitemOpen
  \bibfield  {author} {\bibinfo {author} {\bibfnamefont {A.~Y.}\ \bibnamefont {Kitaev}},\ }\bibfield  {title} {\bibinfo {title} {{Unpaired Majorana fermions in quantum wires}},\ }\href@noop {} {\bibfield  {journal} {\bibinfo  {journal} {Phys. Usp.}\ }\textbf {\bibinfo {volume} {44}},\ \bibinfo {pages} {131} (\bibinfo {year} {2001})}\BibitemShut {NoStop}%
\bibitem [{\citenamefont {Miao}\ \emph {et~al.}(2017)\citenamefont {Miao}, \citenamefont {Jin}, \citenamefont {Zhang},\ and\ \citenamefont {Zhou}}]{miao_exact_2017}%
  \BibitemOpen
  \bibfield  {author} {\bibinfo {author} {\bibfnamefont {J.-J.}\ \bibnamefont {Miao}}, \bibinfo {author} {\bibfnamefont {H.-K.}\ \bibnamefont {Jin}}, \bibinfo {author} {\bibfnamefont {F.-C.}\ \bibnamefont {Zhang}},\ and\ \bibinfo {author} {\bibfnamefont {Y.}~\bibnamefont {Zhou}},\ }\bibfield  {title} {\bibinfo {title} {{Exact Solution for the Interacting Kitaev Chain at the Symmetric Point}},\ }\href {https://doi.org/10.1103/PhysRevLett.118.267701} {\bibfield  {journal} {\bibinfo  {journal} {Phys. Rev. Lett.}\ }\textbf {\bibinfo {volume} {118}},\ \bibinfo {pages} {267701} (\bibinfo {year} {2017})}\BibitemShut {NoStop}%
\bibitem [{\citenamefont {Miao}\ \emph {et~al.}(2018)\citenamefont {Miao}, \citenamefont {Jin}, \citenamefont {Zhang},\ and\ \citenamefont {Zhou}}]{miao_majorana_2018}%
  \BibitemOpen
  \bibfield  {author} {\bibinfo {author} {\bibfnamefont {J.-J.}\ \bibnamefont {Miao}}, \bibinfo {author} {\bibfnamefont {H.-K.}\ \bibnamefont {Jin}}, \bibinfo {author} {\bibfnamefont {F.-C.}\ \bibnamefont {Zhang}},\ and\ \bibinfo {author} {\bibfnamefont {Y.}~\bibnamefont {Zhou}},\ }\bibfield  {title} {\bibinfo {title} {Majorana zero modes and long range edge correlation in interacting {Kitaev} chains: analytic solutions and density-matrix-renormalization-group study},\ }\href {https://doi.org/10.1038/s41598-017-17699-y} {\bibfield  {journal} {\bibinfo  {journal} {Scientific Reports}\ }\textbf {\bibinfo {volume} {8}},\ \bibinfo {pages} {488} (\bibinfo {year} {2018})}\BibitemShut {NoStop}%
\bibitem [{\citenamefont {Lang}\ and\ \citenamefont {B\"{u}chler}(2015)}]{lang_topological_2015}%
  \BibitemOpen
  \bibfield  {author} {\bibinfo {author} {\bibfnamefont {N.}~\bibnamefont {Lang}}\ and\ \bibinfo {author} {\bibfnamefont {H.~P.}\ \bibnamefont {B\"{u}chler}},\ }\bibfield  {title} {\bibinfo {title} {Topological states in a microscopic model of interacting fermions},\ }\href {https://doi.org/10.1103/PhysRevB.92.041118} {\bibfield  {journal} {\bibinfo  {journal} {Phys. Rev. B}\ }\textbf {\bibinfo {volume} {92}},\ \bibinfo {pages} {041118(R)} (\bibinfo {year} {2015})}\BibitemShut {NoStop}%
\bibitem [{\citenamefont {Iemini}\ \emph {et~al.}(2017)\citenamefont {Iemini}, \citenamefont {Mora},\ and\ \citenamefont {Mazza}}]{iemini_topological_2017}%
  \BibitemOpen
  \bibfield  {author} {\bibinfo {author} {\bibfnamefont {F.}~\bibnamefont {Iemini}}, \bibinfo {author} {\bibfnamefont {C.}~\bibnamefont {Mora}},\ and\ \bibinfo {author} {\bibfnamefont {L.}~\bibnamefont {Mazza}},\ }\bibfield  {title} {\bibinfo {title} {{Topological Phases of Parafermions: A Model with Exactly Solvable Ground States}},\ }\href {https://doi.org/10.1103/PhysRevLett.118.170402} {\bibfield  {journal} {\bibinfo  {journal} {Phys. Rev. Lett.}\ }\textbf {\bibinfo {volume} {118}},\ \bibinfo {pages} {170402} (\bibinfo {year} {2017})}\BibitemShut {NoStop}%
\bibitem [{\citenamefont {Weinberg}\ and\ \citenamefont {Bukov}(2017)}]{weinberg_quspin1}%
  \BibitemOpen
  \bibfield  {author} {\bibinfo {author} {\bibfnamefont {P.}~\bibnamefont {Weinberg}}\ and\ \bibinfo {author} {\bibfnamefont {M.}~\bibnamefont {Bukov}},\ }\bibfield  {title} {\bibinfo {title} {{QuSpin: a Python package for dynamics and exact diagonalisation of quantum many body systems part I: spin chains}},\ }\href {https://doi.org/10.21468/SciPostPhys.2.1.003} {\bibfield  {journal} {\bibinfo  {journal} {SciPost Phys.}\ }\textbf {\bibinfo {volume} {2}},\ \bibinfo {pages} {003} (\bibinfo {year} {2017})}\BibitemShut {NoStop}%
\bibitem [{\citenamefont {Weinberg}\ and\ \citenamefont {Bukov}(2019)}]{weinberg_quspin2}%
  \BibitemOpen
  \bibfield  {author} {\bibinfo {author} {\bibfnamefont {P.}~\bibnamefont {Weinberg}}\ and\ \bibinfo {author} {\bibfnamefont {M.}~\bibnamefont {Bukov}},\ }\bibfield  {title} {\bibinfo {title} {{QuSpin: a Python package for dynamics and exact diagonalisation of quantum many body systems. Part II: bosons, fermions and higher spins}},\ }\href {https://doi.org/10.21468/SciPostPhys.7.2.020} {\bibfield  {journal} {\bibinfo  {journal} {SciPost Phys.}\ }\textbf {\bibinfo {volume} {7}},\ \bibinfo {pages} {020} (\bibinfo {year} {2019})}\BibitemShut {NoStop}%
\bibitem [{\citenamefont {Abraham}\ \emph {et~al.}(1997)\citenamefont {Abraham}, \citenamefont {McAlexander}, \citenamefont {Gerton}, \citenamefont {Hulet}, \citenamefont {C\^ot\'e},\ and\ \citenamefont {Dalgarno}}]{Abraham1997}%
  \BibitemOpen
  \bibfield  {author} {\bibinfo {author} {\bibfnamefont {E.~R.~I.}\ \bibnamefont {Abraham}}, \bibinfo {author} {\bibfnamefont {W.~I.}\ \bibnamefont {McAlexander}}, \bibinfo {author} {\bibfnamefont {J.~M.}\ \bibnamefont {Gerton}}, \bibinfo {author} {\bibfnamefont {R.~G.}\ \bibnamefont {Hulet}}, \bibinfo {author} {\bibfnamefont {R.}~\bibnamefont {C\^ot\'e}},\ and\ \bibinfo {author} {\bibfnamefont {A.}~\bibnamefont {Dalgarno}},\ }\bibfield  {title} {\bibinfo {title} {Triplet s-wave resonance in $^{6}\mathrm{Li}$ collisions and scattering lengths of $^{6}\mathrm{Li}$ and $^{7}\mathrm{Li}$},\ }\href {https://doi.org/10.1103/PhysRevA.55.R3299} {\bibfield  {journal} {\bibinfo  {journal} {Phys. Rev. A}\ }\textbf {\bibinfo {volume} {55}},\ \bibinfo {pages} {R3299} (\bibinfo {year} {1997})}\BibitemShut {NoStop}%
\bibitem [{\citenamefont {Bartenstein}\ \emph {et~al.}(2005)\citenamefont {Bartenstein}, \citenamefont {Altmeyer}, \citenamefont {Riedl}, \citenamefont {Geursen}, \citenamefont {Jochim}, \citenamefont {Chin}, \citenamefont {Denschlag}, \citenamefont {Grimm}, \citenamefont {Simoni}, \citenamefont {Tiesinga}, \citenamefont {Williams},\ and\ \citenamefont {Julienne}}]{Bartenstein2005}%
  \BibitemOpen
  \bibfield  {author} {\bibinfo {author} {\bibfnamefont {M.}~\bibnamefont {Bartenstein}}, \bibinfo {author} {\bibfnamefont {A.}~\bibnamefont {Altmeyer}}, \bibinfo {author} {\bibfnamefont {S.}~\bibnamefont {Riedl}}, \bibinfo {author} {\bibfnamefont {R.}~\bibnamefont {Geursen}}, \bibinfo {author} {\bibfnamefont {S.}~\bibnamefont {Jochim}}, \bibinfo {author} {\bibfnamefont {C.}~\bibnamefont {Chin}}, \bibinfo {author} {\bibfnamefont {J.~H.}\ \bibnamefont {Denschlag}}, \bibinfo {author} {\bibfnamefont {R.}~\bibnamefont {Grimm}}, \bibinfo {author} {\bibfnamefont {A.}~\bibnamefont {Simoni}}, \bibinfo {author} {\bibfnamefont {E.}~\bibnamefont {Tiesinga}}, \bibinfo {author} {\bibfnamefont {C.~J.}\ \bibnamefont {Williams}},\ and\ \bibinfo {author} {\bibfnamefont {P.~S.}\ \bibnamefont {Julienne}},\ }\bibfield  {title} {\bibinfo {title} {{Precise Determination of $^{6}\mathrm{Li}$ Cold Collision Parameters by Radio-Frequency Spectroscopy on Weakly Bound Molecules}},\ }\href {https://doi.org/10.1103/PhysRevLett.94.103201}
  {\bibfield  {journal} {\bibinfo  {journal} {Phys. Rev. Lett.}\ }\textbf {\bibinfo {volume} {94}},\ \bibinfo {pages} {103201} (\bibinfo {year} {2005})}\BibitemShut {NoStop}%
\bibitem [{\citenamefont {Fukuhara}\ \emph {et~al.}(2007)\citenamefont {Fukuhara}, \citenamefont {Takasu}, \citenamefont {Kumakura},\ and\ \citenamefont {Takahashi}}]{Fukuhara2007}%
  \BibitemOpen
  \bibfield  {author} {\bibinfo {author} {\bibfnamefont {T.}~\bibnamefont {Fukuhara}}, \bibinfo {author} {\bibfnamefont {Y.}~\bibnamefont {Takasu}}, \bibinfo {author} {\bibfnamefont {M.}~\bibnamefont {Kumakura}},\ and\ \bibinfo {author} {\bibfnamefont {Y.}~\bibnamefont {Takahashi}},\ }\bibfield  {title} {\bibinfo {title} {{Degenerate Fermi Gases of Ytterbium}},\ }\href {https://doi.org/10.1103/PhysRevLett.98.030401} {\bibfield  {journal} {\bibinfo  {journal} {Phys. Rev. Lett.}\ }\textbf {\bibinfo {volume} {98}},\ \bibinfo {pages} {030401} (\bibinfo {year} {2007})}\BibitemShut {NoStop}%
\bibitem [{\citenamefont {Ottenstein}\ \emph {et~al.}(2008)\citenamefont {Ottenstein}, \citenamefont {Lompe}, \citenamefont {Kohnen}, \citenamefont {Wenz},\ and\ \citenamefont {Jochim}}]{Ottenstein2008}%
  \BibitemOpen
  \bibfield  {author} {\bibinfo {author} {\bibfnamefont {T.~B.}\ \bibnamefont {Ottenstein}}, \bibinfo {author} {\bibfnamefont {T.}~\bibnamefont {Lompe}}, \bibinfo {author} {\bibfnamefont {M.}~\bibnamefont {Kohnen}}, \bibinfo {author} {\bibfnamefont {A.~N.}\ \bibnamefont {Wenz}},\ and\ \bibinfo {author} {\bibfnamefont {S.}~\bibnamefont {Jochim}},\ }\bibfield  {title} {\bibinfo {title} {{Collisional Stability of a Three-Component Degenerate Fermi Gas}},\ }\href {https://doi.org/10.1103/PhysRevLett.101.203202} {\bibfield  {journal} {\bibinfo  {journal} {Phys. Rev. Lett.}\ }\textbf {\bibinfo {volume} {101}},\ \bibinfo {pages} {203202} (\bibinfo {year} {2008})}\BibitemShut {NoStop}%
\bibitem [{\citenamefont {Huckans}\ \emph {et~al.}(2009)\citenamefont {Huckans}, \citenamefont {Williams}, \citenamefont {Hazlett}, \citenamefont {Stites},\ and\ \citenamefont {O'Hara}}]{Huckans2009}%
  \BibitemOpen
  \bibfield  {author} {\bibinfo {author} {\bibfnamefont {J.~H.}\ \bibnamefont {Huckans}}, \bibinfo {author} {\bibfnamefont {J.~R.}\ \bibnamefont {Williams}}, \bibinfo {author} {\bibfnamefont {E.~L.}\ \bibnamefont {Hazlett}}, \bibinfo {author} {\bibfnamefont {R.~W.}\ \bibnamefont {Stites}},\ and\ \bibinfo {author} {\bibfnamefont {K.~M.}\ \bibnamefont {O'Hara}},\ }\bibfield  {title} {\bibinfo {title} {{Three-Body Recombination in a Three-State Fermi Gas with Widely Tunable Interactions}},\ }\href {https://doi.org/10.1103/PhysRevLett.102.165302} {\bibfield  {journal} {\bibinfo  {journal} {Phys. Rev. Lett.}\ }\textbf {\bibinfo {volume} {102}},\ \bibinfo {pages} {165302} (\bibinfo {year} {2009})}\BibitemShut {NoStop}%
\bibitem [{\citenamefont {Taie}\ \emph {et~al.}(2010)\citenamefont {Taie}, \citenamefont {Takasu}, \citenamefont {Sugawa}, \citenamefont {Yamazaki}, \citenamefont {Tsujimoto}, \citenamefont {Murakami},\ and\ \citenamefont {Takahashi}}]{Taie2010}%
  \BibitemOpen
  \bibfield  {author} {\bibinfo {author} {\bibfnamefont {S.}~\bibnamefont {Taie}}, \bibinfo {author} {\bibfnamefont {Y.}~\bibnamefont {Takasu}}, \bibinfo {author} {\bibfnamefont {S.}~\bibnamefont {Sugawa}}, \bibinfo {author} {\bibfnamefont {R.}~\bibnamefont {Yamazaki}}, \bibinfo {author} {\bibfnamefont {T.}~\bibnamefont {Tsujimoto}}, \bibinfo {author} {\bibfnamefont {R.}~\bibnamefont {Murakami}},\ and\ \bibinfo {author} {\bibfnamefont {Y.}~\bibnamefont {Takahashi}},\ }\bibfield  {title} {\bibinfo {title} {{Realization of a $\mathrm{SU}(2)\ifmmode\times\else\texttimes\fi{}\mathrm{SU}(6)$ System of Fermions in a Cold Atomic Gas}},\ }\href {https://doi.org/10.1103/PhysRevLett.105.190401} {\bibfield  {journal} {\bibinfo  {journal} {Phys. Rev. Lett.}\ }\textbf {\bibinfo {volume} {105}},\ \bibinfo {pages} {190401} (\bibinfo {year} {2010})}\BibitemShut {NoStop}%
\bibitem [{\citenamefont {Taie}\ \emph {et~al.}(2012)\citenamefont {Taie}, \citenamefont {Yamazaki}, \citenamefont {Sugawa},\ and\ \citenamefont {Takahashi}}]{Taie2012}%
  \BibitemOpen
  \bibfield  {author} {\bibinfo {author} {\bibfnamefont {S.}~\bibnamefont {Taie}}, \bibinfo {author} {\bibfnamefont {R.}~\bibnamefont {Yamazaki}}, \bibinfo {author} {\bibfnamefont {S.}~\bibnamefont {Sugawa}},\ and\ \bibinfo {author} {\bibfnamefont {Y.}~\bibnamefont {Takahashi}},\ }\bibfield  {title} {\bibinfo {title} {{An SU(6) Mott insulator of an atomic Fermi gas realized by large-spin Pomeranchuk cooling}},\ }\href {https://doi.org/10.1038/nphys2430} {\bibfield  {journal} {\bibinfo  {journal} {Nat. Phys.}\ }\textbf {\bibinfo {volume} {8}},\ \bibinfo {pages} {825} (\bibinfo {year} {2012})}\BibitemShut {NoStop}%
\bibitem [{\citenamefont {DeSalvo}\ \emph {et~al.}(2010)\citenamefont {DeSalvo}, \citenamefont {Yan}, \citenamefont {Mickelson}, \citenamefont {Martinez~de Escobar},\ and\ \citenamefont {Killian}}]{Desalvo2010}%
  \BibitemOpen
  \bibfield  {author} {\bibinfo {author} {\bibfnamefont {B.~J.}\ \bibnamefont {DeSalvo}}, \bibinfo {author} {\bibfnamefont {M.}~\bibnamefont {Yan}}, \bibinfo {author} {\bibfnamefont {P.~G.}\ \bibnamefont {Mickelson}}, \bibinfo {author} {\bibfnamefont {Y.~N.}\ \bibnamefont {Martinez~de Escobar}},\ and\ \bibinfo {author} {\bibfnamefont {T.~C.}\ \bibnamefont {Killian}},\ }\bibfield  {title} {\bibinfo {title} {{Degenerate Fermi Gas of $^{87}\mathrm{Sr}$}},\ }\href {https://doi.org/10.1103/PhysRevLett.105.030402} {\bibfield  {journal} {\bibinfo  {journal} {Phys. Rev. Lett.}\ }\textbf {\bibinfo {volume} {105}},\ \bibinfo {pages} {030402} (\bibinfo {year} {2010})}\BibitemShut {NoStop}%
\bibitem [{\citenamefont {Lewenstein}\ \emph {et~al.}(2012)\citenamefont {Lewenstein}, \citenamefont {Sanpera},\ and\ \citenamefont {Ahufinger}}]{Lewenstein2012}%
  \BibitemOpen
  \bibfield  {author} {\bibinfo {author} {\bibfnamefont {M.}~\bibnamefont {Lewenstein}}, \bibinfo {author} {\bibfnamefont {A.}~\bibnamefont {Sanpera}},\ and\ \bibinfo {author} {\bibfnamefont {V.}~\bibnamefont {Ahufinger}},\ }\href {https://doi.org/10.1093/acprof:oso/9780199573127.001.0001} {\emph {\bibinfo {title} {{Ultracold Atoms in Optical Lattices: Simulating Quantum Many-Body Systems}}}}\ (\bibinfo  {publisher} {Oxford University Press, Oxford},\ \bibinfo {year} {2012})\BibitemShut {NoStop}%
\bibitem [{\citenamefont {Scazza}\ \emph {et~al.}(2014)\citenamefont {Scazza}, \citenamefont {Hofrichter}, \citenamefont {H{\"o}fer}, \citenamefont {De~Groot}, \citenamefont {Bloch},\ and\ \citenamefont {F{\"o}lling}}]{Scazza2014}%
  \BibitemOpen
  \bibfield  {author} {\bibinfo {author} {\bibfnamefont {F.}~\bibnamefont {Scazza}}, \bibinfo {author} {\bibfnamefont {C.}~\bibnamefont {Hofrichter}}, \bibinfo {author} {\bibfnamefont {M.}~\bibnamefont {H{\"o}fer}}, \bibinfo {author} {\bibfnamefont {P.~C.}\ \bibnamefont {De~Groot}}, \bibinfo {author} {\bibfnamefont {I.}~\bibnamefont {Bloch}},\ and\ \bibinfo {author} {\bibfnamefont {S.}~\bibnamefont {F{\"o}lling}},\ }\bibfield  {title} {\bibinfo {title} {{Observation of two-orbital spin-exchange interactions with ultracold SU($N$)-symmetric fermions}},\ }\href {https://doi.org/10.1038/nphys3061} {\bibfield  {journal} {\bibinfo  {journal} {Nat. Phys.}\ }\textbf {\bibinfo {volume} {10}},\ \bibinfo {pages} {779} (\bibinfo {year} {2014})}\BibitemShut {NoStop}%
\bibitem [{\citenamefont {Zhang}\ \emph {et~al.}(2014)\citenamefont {Zhang}, \citenamefont {Bishof}, \citenamefont {Bromley}, \citenamefont {Kraus}, \citenamefont {Safronova}, \citenamefont {Zoller}, \citenamefont {Rey},\ and\ \citenamefont {Ye}}]{Zhang2014}%
  \BibitemOpen
  \bibfield  {author} {\bibinfo {author} {\bibfnamefont {X.}~\bibnamefont {Zhang}}, \bibinfo {author} {\bibfnamefont {M.}~\bibnamefont {Bishof}}, \bibinfo {author} {\bibfnamefont {S.~L.}\ \bibnamefont {Bromley}}, \bibinfo {author} {\bibfnamefont {C.~V.}\ \bibnamefont {Kraus}}, \bibinfo {author} {\bibfnamefont {M.~S.}\ \bibnamefont {Safronova}}, \bibinfo {author} {\bibfnamefont {P.}~\bibnamefont {Zoller}}, \bibinfo {author} {\bibfnamefont {A.~M.}\ \bibnamefont {Rey}},\ and\ \bibinfo {author} {\bibfnamefont {J.}~\bibnamefont {Ye}},\ }\bibfield  {title} {\bibinfo {title} {{Spectroscopic observation of SU($N$)-symmetric interactions in Sr orbital magnetism}},\ }\href {https://doi.org/10.1126/science.1254978} {\bibfield  {journal} {\bibinfo  {journal} {Science}\ }\textbf {\bibinfo {volume} {345}},\ \bibinfo {pages} {1467} (\bibinfo {year} {2014})}\BibitemShut {NoStop}%
\bibitem [{\citenamefont {Cazalilla}\ and\ \citenamefont {Rey}(2014)}]{Cazalilla2014}%
  \BibitemOpen
  \bibfield  {author} {\bibinfo {author} {\bibfnamefont {M.~A.}\ \bibnamefont {Cazalilla}}\ and\ \bibinfo {author} {\bibfnamefont {A.~M.}\ \bibnamefont {Rey}},\ }\bibfield  {title} {\bibinfo {title} {{Ultracold Fermi gases with emergent SU($N$) symmetry}},\ }\href {https://doi.org/10.1088/0034-4885/77/12/124401} {\bibfield  {journal} {\bibinfo  {journal} {Rep. Prog. Phys.}\ }\textbf {\bibinfo {volume} {77}},\ \bibinfo {pages} {124401} (\bibinfo {year} {2014})}\BibitemShut {NoStop}%
\bibitem [{\citenamefont {Pagano}\ \emph {et~al.}(2014)\citenamefont {Pagano}, \citenamefont {Mancini}, \citenamefont {Cappellini}, \citenamefont {Lombardi}, \citenamefont {Sch{\"a}fer}, \citenamefont {Hu}, \citenamefont {Liu}, \citenamefont {Catani}, \citenamefont {Sias}, \citenamefont {Inguscio},\ and\ \citenamefont {Fallani}}]{Pagano2014}%
  \BibitemOpen
  \bibfield  {author} {\bibinfo {author} {\bibfnamefont {G.}~\bibnamefont {Pagano}}, \bibinfo {author} {\bibfnamefont {M.}~\bibnamefont {Mancini}}, \bibinfo {author} {\bibfnamefont {G.}~\bibnamefont {Cappellini}}, \bibinfo {author} {\bibfnamefont {P.}~\bibnamefont {Lombardi}}, \bibinfo {author} {\bibfnamefont {F.}~\bibnamefont {Sch{\"a}fer}}, \bibinfo {author} {\bibfnamefont {H.}~\bibnamefont {Hu}}, \bibinfo {author} {\bibfnamefont {X.-J.}\ \bibnamefont {Liu}}, \bibinfo {author} {\bibfnamefont {J.}~\bibnamefont {Catani}}, \bibinfo {author} {\bibfnamefont {C.}~\bibnamefont {Sias}}, \bibinfo {author} {\bibfnamefont {M.}~\bibnamefont {Inguscio}},\ and\ \bibinfo {author} {\bibfnamefont {L.}~\bibnamefont {Fallani}},\ }\bibfield  {title} {\bibinfo {title} {A one-dimensional liquid of fermions with tunable spin},\ }\href {https://doi.org/10.1038/nphys2878} {\bibfield  {journal} {\bibinfo  {journal} {Nat. Phys.}\ }\textbf {\bibinfo {volume} {10}},\ \bibinfo {pages} {198} (\bibinfo {year} {2014})}\BibitemShut
  {NoStop}%
\bibitem [{\citenamefont {Hofrichter}\ \emph {et~al.}(2016)\citenamefont {Hofrichter}, \citenamefont {Riegger}, \citenamefont {Scazza}, \citenamefont {H\"ofer}, \citenamefont {Fernandes}, \citenamefont {Bloch},\ and\ \citenamefont {F\"olling}}]{Hofrichter2016}%
  \BibitemOpen
  \bibfield  {author} {\bibinfo {author} {\bibfnamefont {C.}~\bibnamefont {Hofrichter}}, \bibinfo {author} {\bibfnamefont {L.}~\bibnamefont {Riegger}}, \bibinfo {author} {\bibfnamefont {F.}~\bibnamefont {Scazza}}, \bibinfo {author} {\bibfnamefont {M.}~\bibnamefont {H\"ofer}}, \bibinfo {author} {\bibfnamefont {D.~R.}\ \bibnamefont {Fernandes}}, \bibinfo {author} {\bibfnamefont {I.}~\bibnamefont {Bloch}},\ and\ \bibinfo {author} {\bibfnamefont {S.}~\bibnamefont {F\"olling}},\ }\bibfield  {title} {\bibinfo {title} {{Direct Probing of the Mott Crossover in the $\mathrm{SU}(N)$ Fermi-Hubbard Model}},\ }\href {https://doi.org/10.1103/PhysRevX.6.021030} {\bibfield  {journal} {\bibinfo  {journal} {Phys. Rev. X}\ }\textbf {\bibinfo {volume} {6}},\ \bibinfo {pages} {021030} (\bibinfo {year} {2016})}\BibitemShut {NoStop}%
\bibitem [{\citenamefont {Capponi}\ \emph {et~al.}(2008)\citenamefont {Capponi}, \citenamefont {Roux}, \citenamefont {Lecheminant}, \citenamefont {Azaria}, \citenamefont {Boulat},\ and\ \citenamefont {White}}]{oned2}%
  \BibitemOpen
  \bibfield  {author} {\bibinfo {author} {\bibfnamefont {S.}~\bibnamefont {Capponi}}, \bibinfo {author} {\bibfnamefont {G.}~\bibnamefont {Roux}}, \bibinfo {author} {\bibfnamefont {P.}~\bibnamefont {Lecheminant}}, \bibinfo {author} {\bibfnamefont {P.}~\bibnamefont {Azaria}}, \bibinfo {author} {\bibfnamefont {E.}~\bibnamefont {Boulat}},\ and\ \bibinfo {author} {\bibfnamefont {S.~R.}\ \bibnamefont {White}},\ }\bibfield  {title} {\bibinfo {title} {Molecular superfluid phase in systems of one-dimensional multicomponent fermionic cold atoms},\ }\href {https://doi.org/10.1103/PhysRevA.77.013624} {\bibfield  {journal} {\bibinfo  {journal} {Phys. Rev. A}\ }\textbf {\bibinfo {volume} {77}},\ \bibinfo {pages} {013624} (\bibinfo {year} {2008})}\BibitemShut {NoStop}%
\bibitem [{\citenamefont {Lecheminant}\ \emph {et~al.}(2005)\citenamefont {Lecheminant}, \citenamefont {Boulat},\ and\ \citenamefont {Azaria}}]{oned1}%
  \BibitemOpen
  \bibfield  {author} {\bibinfo {author} {\bibfnamefont {P.}~\bibnamefont {Lecheminant}}, \bibinfo {author} {\bibfnamefont {E.}~\bibnamefont {Boulat}},\ and\ \bibinfo {author} {\bibfnamefont {P.}~\bibnamefont {Azaria}},\ }\bibfield  {title} {\bibinfo {title} {{Confinement and Superfluidity in One-Dimensional Degenerate Fermionic Cold Atoms}},\ }\href {https://doi.org/10.1103/PhysRevLett.95.240402} {\bibfield  {journal} {\bibinfo  {journal} {Phys. Rev. Lett.}\ }\textbf {\bibinfo {volume} {95}},\ \bibinfo {pages} {240402} (\bibinfo {year} {2005})}\BibitemShut {NoStop}%
\bibitem [{\citenamefont {Capponi}\ \emph {et~al.}(2007)\citenamefont {Capponi}, \citenamefont {Roux}, \citenamefont {Azaria}, \citenamefont {Boulat},\ and\ \citenamefont {Lecheminant}}]{oned3}%
  \BibitemOpen
  \bibfield  {author} {\bibinfo {author} {\bibfnamefont {S.}~\bibnamefont {Capponi}}, \bibinfo {author} {\bibfnamefont {G.}~\bibnamefont {Roux}}, \bibinfo {author} {\bibfnamefont {P.}~\bibnamefont {Azaria}}, \bibinfo {author} {\bibfnamefont {E.}~\bibnamefont {Boulat}},\ and\ \bibinfo {author} {\bibfnamefont {P.}~\bibnamefont {Lecheminant}},\ }\bibfield  {title} {\bibinfo {title} {{Confinement versus deconfinement of Cooper pairs in one-dimensional spin-$3/2$ fermionic cold atoms}},\ }\href {https://doi.org/10.1103/PhysRevB.75.100503} {\bibfield  {journal} {\bibinfo  {journal} {Phys. Rev. B}\ }\textbf {\bibinfo {volume} {75}},\ \bibinfo {pages} {100503} (\bibinfo {year} {2007})}\BibitemShut {NoStop}%
\bibitem [{\citenamefont {Roux}\ \emph {et~al.}(2009)\citenamefont {Roux}, \citenamefont {Capponi}, \citenamefont {Lecheminant},\ and\ \citenamefont {Azaria}}]{oned4}%
  \BibitemOpen
  \bibfield  {author} {\bibinfo {author} {\bibfnamefont {G.}~\bibnamefont {Roux}}, \bibinfo {author} {\bibfnamefont {S.}~\bibnamefont {Capponi}}, \bibinfo {author} {\bibfnamefont {P.}~\bibnamefont {Lecheminant}},\ and\ \bibinfo {author} {\bibfnamefont {P.}~\bibnamefont {Azaria}},\ }\bibfield  {title} {\bibinfo {title} {Spin $3/2$ fermions with attractive interactions in a one-dimensional optical lattice: phase diagrams, entanglement entropy, and the effect of the trap},\ }\href@noop {} {\bibfield  {journal} {\bibinfo  {journal} {Eur. Phys. J. B}\ }\textbf {\bibinfo {volume} {68}},\ \bibinfo {pages} {293} (\bibinfo {year} {2009})}\BibitemShut {NoStop}%
\bibitem [{\citenamefont {Soldini}\ \emph {et~al.}(2024)\citenamefont {Soldini}, \citenamefont {Fischer},\ and\ \citenamefont {Neupert}}]{marti_charge_2024}%
  \BibitemOpen
  \bibfield  {author} {\bibinfo {author} {\bibfnamefont {M.~O.}\ \bibnamefont {Soldini}}, \bibinfo {author} {\bibfnamefont {M.~H.}\ \bibnamefont {Fischer}},\ and\ \bibinfo {author} {\bibfnamefont {T.}~\bibnamefont {Neupert}},\ }\bibfield  {title} {\bibinfo {title} {{Charge-$4e$ superconductivity in a Hubbard model}},\ }\href {https://doi.org/10.1103/PhysRevB.109.214509} {\bibfield  {journal} {\bibinfo  {journal} {Phys. Rev. B}\ }\textbf {\bibinfo {volume} {109}},\ \bibinfo {pages} {214509} (\bibinfo {year} {2024})}\BibitemShut {NoStop}%
\bibitem [{\citenamefont {Herland}\ \emph {et~al.}(2010)\citenamefont {Herland}, \citenamefont {Babaev},\ and\ \citenamefont {Sudb\o{}}}]{herland_phase_2010}%
  \BibitemOpen
  \bibfield  {author} {\bibinfo {author} {\bibfnamefont {E.~V.}\ \bibnamefont {Herland}}, \bibinfo {author} {\bibfnamefont {E.}~\bibnamefont {Babaev}},\ and\ \bibinfo {author} {\bibfnamefont {A.}~\bibnamefont {Sudb\o{}}},\ }\bibfield  {title} {\bibinfo {title} {Phase transitions in a three dimensional {$U(1)\ifmmode\times\else\texttimes\fi{}U(1)$} lattice london superconductor: Metallic superfluid and charge-$4e$ superconducting states},\ }\href {https://doi.org/10.1103/PhysRevB.82.134511} {\bibfield  {journal} {\bibinfo  {journal} {Phys. Rev. B}\ }\textbf {\bibinfo {volume} {82}},\ \bibinfo {pages} {134511} (\bibinfo {year} {2010})}\BibitemShut {NoStop}%
\bibitem [{\citenamefont {Jiang}\ \emph {et~al.}(2017)\citenamefont {Jiang}, \citenamefont {Li}, \citenamefont {Kivelson},\ and\ \citenamefont {Yao}}]{4ekivelson}%
  \BibitemOpen
  \bibfield  {author} {\bibinfo {author} {\bibfnamefont {Y.-F.}\ \bibnamefont {Jiang}}, \bibinfo {author} {\bibfnamefont {Z.-X.}\ \bibnamefont {Li}}, \bibinfo {author} {\bibfnamefont {S.~A.}\ \bibnamefont {Kivelson}},\ and\ \bibinfo {author} {\bibfnamefont {H.}~\bibnamefont {Yao}},\ }\bibfield  {title} {\bibinfo {title} {{Charge-$4e$ superconductors: A Majorana quantum Monte Carlo study}},\ }\href {https://doi.org/10.1103/PhysRevB.95.241103} {\bibfield  {journal} {\bibinfo  {journal} {Phys. Rev. B}\ }\textbf {\bibinfo {volume} {95}},\ \bibinfo {pages} {241103} (\bibinfo {year} {2017})}\BibitemShut {NoStop}%
\bibitem [{\citenamefont {Fernandes}\ and\ \citenamefont {Fu}(2021)}]{4efernandes}%
  \BibitemOpen
  \bibfield  {author} {\bibinfo {author} {\bibfnamefont {R.~M.}\ \bibnamefont {Fernandes}}\ and\ \bibinfo {author} {\bibfnamefont {L.}~\bibnamefont {Fu}},\ }\bibfield  {title} {\bibinfo {title} {{Charge-$4e$ Superconductivity from Multicomponent Nematic Pairing: Application to Twisted Bilayer Graphene}},\ }\href {https://doi.org/10.1103/PhysRevLett.127.047001} {\bibfield  {journal} {\bibinfo  {journal} {Phys. Rev. Lett.}\ }\textbf {\bibinfo {volume} {127}},\ \bibinfo {pages} {047001} (\bibinfo {year} {2021})}\BibitemShut {NoStop}%
\bibitem [{\citenamefont {Nazaryan}\ and\ \citenamefont {Fu}(2024)}]{magnonSC}%
  \BibitemOpen
  \bibfield  {author} {\bibinfo {author} {\bibfnamefont {K.~G.}\ \bibnamefont {Nazaryan}}\ and\ \bibinfo {author} {\bibfnamefont {L.}~\bibnamefont {Fu}},\ }\href {https://arxiv.org/abs/2403.14756} {\bibinfo {title} {Magnonic superconductivity}} (\bibinfo {year} {2024}),\ \Eprint {https://arxiv.org/abs/2403.14756} {arXiv:2403.14756 [cond-mat.supr-con]} \BibitemShut {NoStop}%
\bibitem [{\citenamefont {Efimov}(1970{\natexlab{a}})}]{efimov_weakly_1970}%
  \BibitemOpen
  \bibfield  {author} {\bibinfo {author} {\bibfnamefont {V.~N.}\ \bibnamefont {Efimov}},\ }\bibfield  {title} {\bibinfo {title} {{WEAKLY} {BOUND} {STATES} {OF} {THREE} {RESONANTLY} {INTERACTING} {PARTICLES}.},\ }\href {https://www.osti.gov/biblio/4068792} {\bibfield  {journal} {\bibinfo  {journal} {Yadern. Fiz.}\ ,\ \bibinfo {pages} {1080}} (\bibinfo {year} {1970}{\natexlab{a}})}\BibitemShut {NoStop}%
\bibitem [{\citenamefont {Efimov}(1970{\natexlab{b}})}]{efimov_energy_1970}%
  \BibitemOpen
  \bibfield  {author} {\bibinfo {author} {\bibfnamefont {V.}~\bibnamefont {Efimov}},\ }\bibfield  {title} {\bibinfo {title} {Energy levels arising from resonant two-body forces in a three-body system},\ }\href@noop {} {\bibfield  {journal} {\bibinfo  {journal} {Physics Letters B}\ }\textbf {\bibinfo {volume} {33}},\ \bibinfo {pages} {563} (\bibinfo {year} {1970}{\natexlab{b}})}\BibitemShut {NoStop}%
\bibitem [{\citenamefont {Naidon}\ and\ \citenamefont {Endo}(2017)}]{naidon_efimov_2017}%
  \BibitemOpen
  \bibfield  {author} {\bibinfo {author} {\bibfnamefont {P.}~\bibnamefont {Naidon}}\ and\ \bibinfo {author} {\bibfnamefont {S.}~\bibnamefont {Endo}},\ }\bibfield  {title} {\bibinfo {title} {Efimov physics: a review},\ }\href {https://doi.org/10.1088/1361-6633/aa50e8} {\bibfield  {journal} {\bibinfo  {journal} {Reports on Progress in Physics}\ }\textbf {\bibinfo {volume} {80}},\ \bibinfo {pages} {056001} (\bibinfo {year} {2017})}\BibitemShut {NoStop}%
\bibitem [{\citenamefont {Niemann}\ and\ \citenamefont {Hammer}(2012)}]{niemann_pauli_2012}%
  \BibitemOpen
  \bibfield  {author} {\bibinfo {author} {\bibfnamefont {P.}~\bibnamefont {Niemann}}\ and\ \bibinfo {author} {\bibfnamefont {H.-W.}\ \bibnamefont {Hammer}},\ }\bibfield  {title} {\bibinfo {title} {Pauli-blocking effects and cooper triples in three-component fermi gases},\ }\href {https://doi.org/10.1103/PhysRevA.86.013628} {\bibfield  {journal} {\bibinfo  {journal} {Phys. Rev. A}\ }\textbf {\bibinfo {volume} {86}},\ \bibinfo {pages} {013628} (\bibinfo {year} {2012})}\BibitemShut {NoStop}%
\bibitem [{\citenamefont {Tajima}\ \emph {et~al.}(2020)\citenamefont {Tajima}, \citenamefont {Tsutsui}, \citenamefont {Doi},\ and\ \citenamefont {Iida}}]{tajima_cooper_2020}%
  \BibitemOpen
  \bibfield  {author} {\bibinfo {author} {\bibfnamefont {H.}~\bibnamefont {Tajima}}, \bibinfo {author} {\bibfnamefont {S.}~\bibnamefont {Tsutsui}}, \bibinfo {author} {\bibfnamefont {T.~M.}\ \bibnamefont {Doi}},\ and\ \bibinfo {author} {\bibfnamefont {K.}~\bibnamefont {Iida}},\ }\bibfield  {title} {\bibinfo {title} {Cooper {Triples} in {Attractive} {SU}(3) {Fermions} with an {Asymptotic} {Freedom}},\ }\href {http://arxiv.org/abs/2012.03627} {\bibfield  {journal} {\bibinfo  {journal} {arXiv: 2012.03627}\ } (\bibinfo {year} {2020})}\BibitemShut {NoStop}%
\bibitem [{\citenamefont {Akagami}\ \emph {et~al.}(2021)\citenamefont {Akagami}, \citenamefont {Tajima},\ and\ \citenamefont {Iida}}]{akagami_condensation_2021}%
  \BibitemOpen
  \bibfield  {author} {\bibinfo {author} {\bibfnamefont {S.}~\bibnamefont {Akagami}}, \bibinfo {author} {\bibfnamefont {H.}~\bibnamefont {Tajima}},\ and\ \bibinfo {author} {\bibfnamefont {K.}~\bibnamefont {Iida}},\ }\bibfield  {title} {\bibinfo {title} {Condensation of {Cooper} triples},\ }\href {https://doi.org/10.1103/PhysRevA.104.L041302} {\bibfield  {journal} {\bibinfo  {journal} {Phys. Rev. A}\ }\textbf {\bibinfo {volume} {104}},\ \bibinfo {pages} {L041302} (\bibinfo {year} {2021})}\BibitemShut {NoStop}%
\bibitem [{\citenamefont {Tajima}\ \emph {et~al.}(2021)\citenamefont {Tajima}, \citenamefont {Tsutsui}, \citenamefont {Doi},\ and\ \citenamefont {Iida}}]{tajima_three_2021}%
  \BibitemOpen
  \bibfield  {author} {\bibinfo {author} {\bibfnamefont {H.}~\bibnamefont {Tajima}}, \bibinfo {author} {\bibfnamefont {S.}~\bibnamefont {Tsutsui}}, \bibinfo {author} {\bibfnamefont {T.~M.}\ \bibnamefont {Doi}},\ and\ \bibinfo {author} {\bibfnamefont {K.}~\bibnamefont {Iida}},\ }\bibfield  {title} {\bibinfo {title} {Three-body crossover from a {Cooper} triple to a bound trimer state in three-component fermi gases near a triatomic resonance},\ }\href {https://doi.org/10.1103/PhysRevA.104.053328} {\bibfield  {journal} {\bibinfo  {journal} {Phys. Rev. A}\ }\textbf {\bibinfo {volume} {104}},\ \bibinfo {pages} {053328} (\bibinfo {year} {2021})}\BibitemShut {NoStop}%
\bibitem [{\citenamefont {Pérez-Romero}\ \emph {et~al.}(2021)\citenamefont {Pérez-Romero}, \citenamefont {Franco},\ and\ \citenamefont {Silva-Valencia}}]{perez-romero_phase_2021}%
  \BibitemOpen
  \bibfield  {author} {\bibinfo {author} {\bibfnamefont {A.}~\bibnamefont {Pérez-Romero}}, \bibinfo {author} {\bibfnamefont {R.}~\bibnamefont {Franco}},\ and\ \bibinfo {author} {\bibfnamefont {J.}~\bibnamefont {Silva-Valencia}},\ }\bibfield  {title} {\bibinfo {title} {Phase diagram of the {SU}(3) {Fermi}–{Hubbard} model with next-neighbor interactions},\ }\href {https://doi.org/10.1140/epjb/s10051-021-00242-4} {\bibfield  {journal} {\bibinfo  {journal} {The European Physical Journal B}\ }\textbf {\bibinfo {volume} {94}},\ \bibinfo {pages} {229} (\bibinfo {year} {2021})}\BibitemShut {NoStop}%
\bibitem [{\citenamefont {Penson}\ and\ \citenamefont {Kolb}(1986)}]{penson_real-space_1986}%
  \BibitemOpen
  \bibfield  {author} {\bibinfo {author} {\bibfnamefont {K.~A.}\ \bibnamefont {Penson}}\ and\ \bibinfo {author} {\bibfnamefont {M.}~\bibnamefont {Kolb}},\ }\bibfield  {title} {\bibinfo {title} {Real-space pairing in fermion systems},\ }\href {https://doi.org/10.1103/PhysRevB.33.1663} {\bibfield  {journal} {\bibinfo  {journal} {Phys. Rev. B}\ }\textbf {\bibinfo {volume} {33}},\ \bibinfo {pages} {1663} (\bibinfo {year} {1986})}\BibitemShut {NoStop}%
\bibitem [{\citenamefont {Affleck}\ and\ \citenamefont {Marston}(1988)}]{affleck_field-theory_1988}%
  \BibitemOpen
  \bibfield  {author} {\bibinfo {author} {\bibfnamefont {I.}~\bibnamefont {Affleck}}\ and\ \bibinfo {author} {\bibfnamefont {J.~B.}\ \bibnamefont {Marston}},\ }\bibfield  {title} {\bibinfo {title} {Field-theory analysis of a short-range pairing model},\ }\href {https://doi.org/10.1088/0022-3719/21/13/014} {\bibfield  {journal} {\bibinfo  {journal} {J. Phys. C: Solid State Phys.}\ }\textbf {\bibinfo {volume} {21}},\ \bibinfo {pages} {2511} (\bibinfo {year} {1988})}\BibitemShut {NoStop}%
\bibitem [{\citenamefont {Hui}\ and\ \citenamefont {Doniach}(1993)}]{hui_penson_1993}%
  \BibitemOpen
  \bibfield  {author} {\bibinfo {author} {\bibfnamefont {A.}~\bibnamefont {Hui}}\ and\ \bibinfo {author} {\bibfnamefont {S.}~\bibnamefont {Doniach}},\ }\bibfield  {title} {\bibinfo {title} {{Penson-Kolb-Hubbard model: A study of the competition between single-particle and pair hopping in one dimension}},\ }\href {https://doi.org/10.1103/PhysRevB.48.2063} {\bibfield  {journal} {\bibinfo  {journal} {Phys. Rev. B}\ }\textbf {\bibinfo {volume} {48}},\ \bibinfo {pages} {2063} (\bibinfo {year} {1993})}\BibitemShut {NoStop}%
\bibitem [{\citenamefont {Arrachea}\ \emph {et~al.}(1997)\citenamefont {Arrachea}, \citenamefont {Gagliano},\ and\ \citenamefont {Aligia}}]{arrachea_ground-state_1997}%
  \BibitemOpen
  \bibfield  {author} {\bibinfo {author} {\bibfnamefont {L.}~\bibnamefont {Arrachea}}, \bibinfo {author} {\bibfnamefont {E.~R.}\ \bibnamefont {Gagliano}},\ and\ \bibinfo {author} {\bibfnamefont {A.~A.}\ \bibnamefont {Aligia}},\ }\bibfield  {title} {\bibinfo {title} {Ground-state phase diagram of an extended {Hubbard} chain with correlated hopping at half-filling},\ }\href {https://doi.org/10.1103/PhysRevB.55.1173} {\bibfield  {journal} {\bibinfo  {journal} {Phys. Rev. B}\ }\textbf {\bibinfo {volume} {55}},\ \bibinfo {pages} {1173} (\bibinfo {year} {1997})}\BibitemShut {NoStop}%
\bibitem [{\citenamefont {Japaridze}\ \emph {et~al.}(2001)\citenamefont {Japaridze}, \citenamefont {Kampf}, \citenamefont {Sekania}, \citenamefont {Kakashvili},\ and\ \citenamefont {Brune}}]{japaridze_eta_2001}%
  \BibitemOpen
  \bibfield  {author} {\bibinfo {author} {\bibfnamefont {G.~I.}\ \bibnamefont {Japaridze}}, \bibinfo {author} {\bibfnamefont {A.~P.}\ \bibnamefont {Kampf}}, \bibinfo {author} {\bibfnamefont {M.}~\bibnamefont {Sekania}}, \bibinfo {author} {\bibfnamefont {P.}~\bibnamefont {Kakashvili}},\ and\ \bibinfo {author} {\bibfnamefont {P.}~\bibnamefont {Brune}},\ }\bibfield  {title} {\bibinfo {title} {{$\eta$-pairing superconductivity in the Hubbard chain with pair hopping}},\ }\href {https://doi.org/10.1103/PhysRevB.65.014518} {\bibfield  {journal} {\bibinfo  {journal} {Phys. Rev. B}\ }\textbf {\bibinfo {volume} {65}},\ \bibinfo {pages} {014518} (\bibinfo {year} {2001})}\BibitemShut {NoStop}%
\bibitem [{\citenamefont {Dutta}\ \emph {et~al.}(2015)\citenamefont {Dutta}, \citenamefont {Gajda}, \citenamefont {Hauke}, \citenamefont {Lewenstein}, \citenamefont {Lühmann}, \citenamefont {Malomed}, \citenamefont {Sowiński},\ and\ \citenamefont {Zakrzewski}}]{dutta_non-standard_2015}%
  \BibitemOpen
  \bibfield  {author} {\bibinfo {author} {\bibfnamefont {O.}~\bibnamefont {Dutta}}, \bibinfo {author} {\bibfnamefont {M.}~\bibnamefont {Gajda}}, \bibinfo {author} {\bibfnamefont {P.}~\bibnamefont {Hauke}}, \bibinfo {author} {\bibfnamefont {M.}~\bibnamefont {Lewenstein}}, \bibinfo {author} {\bibfnamefont {D.-S.}\ \bibnamefont {Lühmann}}, \bibinfo {author} {\bibfnamefont {B.~A.}\ \bibnamefont {Malomed}}, \bibinfo {author} {\bibfnamefont {T.}~\bibnamefont {Sowiński}},\ and\ \bibinfo {author} {\bibfnamefont {J.}~\bibnamefont {Zakrzewski}},\ }\bibfield  {title} {\bibinfo {title} {Non-standard {Hubbard} models in optical lattices: a review},\ }\href {https://doi.org/10.1088/0034-4885/78/6/066001} {\bibfield  {journal} {\bibinfo  {journal} {Rep. Prog. Phys.}\ }\textbf {\bibinfo {volume} {78}},\ \bibinfo {pages} {066001} (\bibinfo {year} {2015})}\BibitemShut {NoStop}%
\bibitem [{\citenamefont {Sous}\ \emph {et~al.}(2018)\citenamefont {Sous}, \citenamefont {Chakraborty}, \citenamefont {Krems},\ and\ \citenamefont {Berciu}}]{sous_light_2018}%
  \BibitemOpen
  \bibfield  {author} {\bibinfo {author} {\bibfnamefont {J.}~\bibnamefont {Sous}}, \bibinfo {author} {\bibfnamefont {M.}~\bibnamefont {Chakraborty}}, \bibinfo {author} {\bibfnamefont {R.~V.}\ \bibnamefont {Krems}},\ and\ \bibinfo {author} {\bibfnamefont {M.}~\bibnamefont {Berciu}},\ }\bibfield  {title} {\bibinfo {title} {{Light Bipolarons Stabilized by Peierls Electron-Phonon Coupling}},\ }\href {https://doi.org/10.1103/PhysRevLett.121.247001} {\bibfield  {journal} {\bibinfo  {journal} {Phys. Rev. Lett.}\ }\textbf {\bibinfo {volume} {121}},\ \bibinfo {pages} {247001} (\bibinfo {year} {2018})}\BibitemShut {NoStop}%
\bibitem [{\citenamefont {M\"oller}\ \emph {et~al.}(2019)\citenamefont {M\"oller}, \citenamefont {Adolphs},\ and\ \citenamefont {Berciu}}]{moller_magnon_2019}%
  \BibitemOpen
  \bibfield  {author} {\bibinfo {author} {\bibfnamefont {M.~M.}\ \bibnamefont {M\"oller}}, \bibinfo {author} {\bibfnamefont {C.~P.~J.}\ \bibnamefont {Adolphs}},\ and\ \bibinfo {author} {\bibfnamefont {M.}~\bibnamefont {Berciu}},\ }\bibfield  {title} {\bibinfo {title} {Magnon-mediated attraction between two holes doped in a ${\mathrm{cuo}}_{2}$ layer},\ }\href {https://doi.org/10.1103/PhysRevB.100.165118} {\bibfield  {journal} {\bibinfo  {journal} {Phys. Rev. B}\ }\textbf {\bibinfo {volume} {100}},\ \bibinfo {pages} {165118} (\bibinfo {year} {2019})}\BibitemShut {NoStop}%
\bibitem [{\citenamefont {Sous}\ and\ \citenamefont {Pretko}(2020)}]{sous_fracton_2020}%
  \BibitemOpen
  \bibfield  {author} {\bibinfo {author} {\bibfnamefont {J.}~\bibnamefont {Sous}}\ and\ \bibinfo {author} {\bibfnamefont {M.}~\bibnamefont {Pretko}},\ }\bibfield  {title} {\bibinfo {title} {Fractons from polarons},\ }\href {https://doi.org/10.1103/PhysRevB.102.214437} {\bibfield  {journal} {\bibinfo  {journal} {Phys. Rev. B}\ }\textbf {\bibinfo {volume} {102}},\ \bibinfo {pages} {214437} (\bibinfo {year} {2020})}\BibitemShut {NoStop}%
\bibitem [{\citenamefont {Gotta}\ \emph {et~al.}(2021)\citenamefont {Gotta}, \citenamefont {Mazza}, \citenamefont {Simon},\ and\ \citenamefont {Roux}}]{gotta_two-fluid_2021}%
  \BibitemOpen
  \bibfield  {author} {\bibinfo {author} {\bibfnamefont {L.}~\bibnamefont {Gotta}}, \bibinfo {author} {\bibfnamefont {L.}~\bibnamefont {Mazza}}, \bibinfo {author} {\bibfnamefont {P.}~\bibnamefont {Simon}},\ and\ \bibinfo {author} {\bibfnamefont {G.}~\bibnamefont {Roux}},\ }\bibfield  {title} {\bibinfo {title} {Two-fluid coexistence and phase separation in a one-dimensional model with pair hopping and density interactions},\ }\href {https://doi.org/10.1103/PhysRevB.104.094521} {\bibfield  {journal} {\bibinfo  {journal} {Phys. Rev. B}\ }\textbf {\bibinfo {volume} {104}},\ \bibinfo {pages} {094521} (\bibinfo {year} {2021})}\BibitemShut {NoStop}%
\bibitem [{\citenamefont {Zhao}\ \emph {et~al.}(2007)\citenamefont {Zhao}, \citenamefont {Ueda},\ and\ \citenamefont {Wang}}]{zhao_insulating_2007}%
  \BibitemOpen
  \bibfield  {author} {\bibinfo {author} {\bibfnamefont {J.}~\bibnamefont {Zhao}}, \bibinfo {author} {\bibfnamefont {K.}~\bibnamefont {Ueda}},\ and\ \bibinfo {author} {\bibfnamefont {X.}~\bibnamefont {Wang}},\ }\bibfield  {title} {\bibinfo {title} {{Insulating charge density wave for a half-filled SU($N$) Hubbard model with an attractive on-site interaction in one dimension}},\ }\href {https://doi.org/10.1143/JPSJ.76.114711} {\bibfield  {journal} {\bibinfo  {journal} {J. Phys. Jpn.}\ }\textbf {\bibinfo {volume} {76}},\ \bibinfo {pages} {114711} (\bibinfo {year} {2007})}\BibitemShut {NoStop}%
\bibitem [{\citenamefont {Nonne}\ \emph {et~al.}(2011)\citenamefont {Nonne}, \citenamefont {Lecheminant}, \citenamefont {Capponi}, \citenamefont {Roux},\ and\ \citenamefont {Boulat}}]{nonne_competing_2011}%
  \BibitemOpen
  \bibfield  {author} {\bibinfo {author} {\bibfnamefont {H.}~\bibnamefont {Nonne}}, \bibinfo {author} {\bibfnamefont {P.}~\bibnamefont {Lecheminant}}, \bibinfo {author} {\bibfnamefont {S.}~\bibnamefont {Capponi}}, \bibinfo {author} {\bibfnamefont {G.}~\bibnamefont {Roux}},\ and\ \bibinfo {author} {\bibfnamefont {E.}~\bibnamefont {Boulat}},\ }\bibfield  {title} {\bibinfo {title} {{Competing orders in one-dimensional half-filled multicomponent fermionic cold atoms: The Haldane-charge conjecture}},\ }\href {https://doi.org/10.1103/PhysRevB.84.125123} {\bibfield  {journal} {\bibinfo  {journal} {Phys. Rev. B}\ }\textbf {\bibinfo {volume} {84}},\ \bibinfo {pages} {125123} (\bibinfo {year} {2011})}\BibitemShut {NoStop}%
\bibitem [{\citenamefont {Capponi}\ \emph {et~al.}(2016)\citenamefont {Capponi}, \citenamefont {Lecheminant},\ and\ \citenamefont {Totsuka}}]{capponi_2016}%
  \BibitemOpen
  \bibfield  {author} {\bibinfo {author} {\bibfnamefont {S.}~\bibnamefont {Capponi}}, \bibinfo {author} {\bibfnamefont {P.}~\bibnamefont {Lecheminant}},\ and\ \bibinfo {author} {\bibfnamefont {K.}~\bibnamefont {Totsuka}},\ }\bibfield  {title} {\bibinfo {title} {{Phases of one-dimensional {SU}({$N$}) cold atomic {Fermi} gases—{From} molecular {Luttinger} liquids to topological phases}},\ }\href {https://doi.org/10.1016/j.aop.2016.01.011} {\bibfield  {journal} {\bibinfo  {journal} {Ann. Phys.}\ }\textbf {\bibinfo {volume} {367}},\ \bibinfo {pages} {50} (\bibinfo {year} {2016})}\BibitemShut {NoStop}%
\bibitem [{Note1()}]{Note1}%
  \BibitemOpen
  \bibinfo {note} {Without loss of generality, we can assume that $t_1 \geq 0$; if $t_1 < 0$, we can change its sign by flipping the sign of $\protect \hat {c}_{j,\alpha }$ at every other site, i.e., $\protect \hat {c}_{j,\alpha } \to (-1)^j \protect \hat {c}_{j,\alpha }$. Note that this transformation does not change the signs of the other parameters of the model.}\BibitemShut {Stop}%
\bibitem [{Note2()}]{Note2}%
  \BibitemOpen
  \bibinfo {note} {Note that these terms can be rewritten as \\$\protect \hat {H}_{U} =U \DOTSB \sum@ \slimits@ _{j=1}^{L} \DOTSB \sum@ \slimits@ _{1\leq \alpha <\beta \leq 3}\left (\protect \hat {n}_{j,\alpha }-\protect \frac {1}{2}\right )\left (\protect \hat {n}_{j,\beta }-\protect \frac {1}{2}\right )+\protect \text {const.}$ and $\protect \hat {H}_{V} =V \DOTSB \sum@ \slimits@ _{j=1}^{L-1} \DOTSB \sum@ \slimits@ _{\alpha ,\beta = 1}^3 \left (\protect \hat {n}_{j,\alpha }-\protect \frac {1}{2}\right )\left (\protect \hat {n}_{j+1,\beta }-\protect \frac {1}{2}\right )$.}\BibitemShut {Stop}%
\bibitem [{Note3()}]{Note3}%
  \BibitemOpen
  \bibinfo {note} {This $\eta $-operator generalizes the one introduced in Ref.~\cite {yang_1989_eta}. In the SU(2) Hubbard model, the $\eta $-operator is defined as $\protect \hat {\protect \tilde {\eta }}_2^\dagger =\DOTSB \sum@ \slimits@ _{j=1}^L (-1)^j\protect \hat {c}^{\dagger }_{j,1}\protect \hat {c}^{\dagger }_{j,2}$. One might naively attempt to generalize it to the SU(3) case by defining $\protect \hat {\protect \tilde {\eta }}_3^\dagger =\DOTSB \sum@ \slimits@ _{j=1}^L (-1)^j \protect \hat {\eta }_{j}^\dagger $, but this approach fails because $\protect \hat {\protect \tilde {\eta }}_3^\dagger $ squares to zero. To circumvent this issue, we introduced the Jordan-Wigner string $\protect \hat {U}_{j-1}$ so that $({\protect \hat \eta }^\dagger )^2$ remains nonzero.}\BibitemShut {Stop}%
\bibitem [{\citenamefont {Vafek}\ \emph {et~al.}(2017)\citenamefont {Vafek}, \citenamefont {Regnault},\ and\ \citenamefont {Bernevig}}]{vafek_entanglement_2017}%
  \BibitemOpen
  \bibfield  {author} {\bibinfo {author} {\bibfnamefont {O.}~\bibnamefont {Vafek}}, \bibinfo {author} {\bibfnamefont {N.}~\bibnamefont {Regnault}},\ and\ \bibinfo {author} {\bibfnamefont {B.~A.}\ \bibnamefont {Bernevig}},\ }\bibfield  {title} {\bibinfo {title} {Entanglement of {Exact} {Excited} {Eigenstates} of the {Hubbard} {Model} in {Arbitrary} {Dimension}},\ }\href {https://doi.org/10.21468/SciPostPhys.3.6.043} {\bibfield  {journal} {\bibinfo  {journal} {SciPost Phys.}\ }\textbf {\bibinfo {volume} {3}},\ \bibinfo {pages} {043} (\bibinfo {year} {2017})}\BibitemShut {NoStop}%
\bibitem [{\citenamefont {Vedral}(2004)}]{vedral_high-temperature_2004}%
  \BibitemOpen
  \bibfield  {author} {\bibinfo {author} {\bibfnamefont {V.}~\bibnamefont {Vedral}},\ }\bibfield  {title} {\bibinfo {title} {High-temperature macroscopic entanglement},\ }\href {https://doi.org/10.1088/1367-2630/6/1/102} {\bibfield  {journal} {\bibinfo  {journal} {New J. Phys.}\ }\textbf {\bibinfo {volume} {6}},\ \bibinfo {pages} {102} (\bibinfo {year} {2004})}\BibitemShut {NoStop}%
\bibitem [{\citenamefont {Fan}\ and\ \citenamefont {Lloyd}(2005)}]{fan_entanglement_2005}%
  \BibitemOpen
  \bibfield  {author} {\bibinfo {author} {\bibfnamefont {H.}~\bibnamefont {Fan}}\ and\ \bibinfo {author} {\bibfnamefont {S.}~\bibnamefont {Lloyd}},\ }\bibfield  {title} {\bibinfo {title} {Entanglement and off-diagonal long-range order of an {$\eta$}-pairing state},\ }\href {https://doi.org/10.1088/0305-4470/38/23/014} {\bibfield  {journal} {\bibinfo  {journal} {J. Phys. A: Math. Gen.}\ }\textbf {\bibinfo {volume} {38}},\ \bibinfo {pages} {5285} (\bibinfo {year} {2005})}\BibitemShut {NoStop}%
\bibitem [{\citenamefont {Efetov}\ and\ \citenamefont {Larkin}(1975)}]{efetov_correlation_1975}%
  \BibitemOpen
  \bibfield  {author} {\bibinfo {author} {\bibfnamefont {K.}~\bibnamefont {Efetov}}\ and\ \bibinfo {author} {\bibfnamefont {A.}~\bibnamefont {Larkin}},\ }\bibfield  {title} {\bibinfo {title} {Correlation functions in one-dimensional systems with a strong interaction},\ }\href@noop {} {\bibfield  {journal} {\bibinfo  {journal} {Soviet Phys. JETP}\ }\textbf {\bibinfo {volume} {42}},\ \bibinfo {pages} {390} (\bibinfo {year} {1975})}\BibitemShut {NoStop}%
\bibitem [{\citenamefont {Emery}(1976)}]{emery_theory_1976}%
  \BibitemOpen
  \bibfield  {author} {\bibinfo {author} {\bibfnamefont {V.~J.}\ \bibnamefont {Emery}},\ }\bibfield  {title} {\bibinfo {title} {Theory of the quasi-one-dimensional electron gas with strong "on-site" interactions},\ }\href {https://doi.org/10.1103/PhysRevB.14.2989} {\bibfield  {journal} {\bibinfo  {journal} {Phys. Rev. B}\ }\textbf {\bibinfo {volume} {14}},\ \bibinfo {pages} {2989} (\bibinfo {year} {1976})}\BibitemShut {NoStop}%
\bibitem [{\citenamefont {Giamarchi}(2003)}]{giamarchi_quantum_2003}%
  \BibitemOpen
  \bibfield  {author} {\bibinfo {author} {\bibfnamefont {T.}~\bibnamefont {Giamarchi}},\ }\href {https://doi.org/10.1093/acprof:oso/9780198525004.001.0001} {\emph {\bibinfo {title} {Quantum {Physics} in {One} {Dimension}}}}\ (\bibinfo  {publisher} {Oxford University Press},\ \bibinfo {year} {2003})\BibitemShut {NoStop}%
\bibitem [{Note4()}]{Note4}%
  \BibitemOpen
  \bibinfo {note} {Here, phase separation refers to a state in which particles tend to aggregate due to attractive interactions, resulting in two distinct regions: one particle-rich and the other particle-poor.}\BibitemShut {Stop}%
\bibitem [{Note5()}]{Note5}%
  \BibitemOpen
  \bibinfo {note} {This ground state is stable against certain types of perturbations. For example, the $\eta $-clustering state is a ground state of local Hamiltonians $\protect \hat {h}_{j,j+1} = \protect \frac {1}{2}(\protect \hat {\eta }^{\dagger }_{j}\protect \hat {\eta }_{j+1}+\protect \text {h.c.)} -\protect \frac {1}{9} (\protect \hat {n}_j-\protect \frac {3}{2}) (\protect \hat {n}_{j+1}-\protect \frac {3}{2})$~\cite {yoshida_exact_2022}. Thus, even if we add $\protect \hat {h}_{j,j+1}$ terms to our Hamiltonian, $\eta $-clustering state is expected to remain a ground state. The first term of $\protect \hat {h}_{j,j+1}$ describes the three-body hopping term. The second term represents the next-nearest attractive interactions, which can be absorbed into $\protect \hat {H}_V$.}\BibitemShut {Stop}%
\bibitem [{\citenamefont {Sala}\ \emph {et~al.}(2020)\citenamefont {Sala}, \citenamefont {Rakovszky}, \citenamefont {Verresen}, \citenamefont {Knap},\ and\ \citenamefont {Pollmann}}]{sala_ergodicity_2020}%
  \BibitemOpen
  \bibfield  {author} {\bibinfo {author} {\bibfnamefont {P.}~\bibnamefont {Sala}}, \bibinfo {author} {\bibfnamefont {T.}~\bibnamefont {Rakovszky}}, \bibinfo {author} {\bibfnamefont {R.}~\bibnamefont {Verresen}}, \bibinfo {author} {\bibfnamefont {M.}~\bibnamefont {Knap}},\ and\ \bibinfo {author} {\bibfnamefont {F.}~\bibnamefont {Pollmann}},\ }\bibfield  {title} {\bibinfo {title} {Ergodicity {Breaking} {Arising} from {Hilbert} {Space} {Fragmentation} in {Dipole}-{Conserving} {Hamiltonians}},\ }\href {https://doi.org/10.1103/PhysRevX.10.011047} {\bibfield  {journal} {\bibinfo  {journal} {Phys. Rev. X}\ }\textbf {\bibinfo {volume} {10}},\ \bibinfo {pages} {011047} (\bibinfo {year} {2020})}\BibitemShut {NoStop}%
\bibitem [{\citenamefont {Khemani}\ \emph {et~al.}(2020)\citenamefont {Khemani}, \citenamefont {Hermele},\ and\ \citenamefont {Nandkishore}}]{khemani_localization_2020}%
  \BibitemOpen
  \bibfield  {author} {\bibinfo {author} {\bibfnamefont {V.}~\bibnamefont {Khemani}}, \bibinfo {author} {\bibfnamefont {M.}~\bibnamefont {Hermele}},\ and\ \bibinfo {author} {\bibfnamefont {R.}~\bibnamefont {Nandkishore}},\ }\bibfield  {title} {\bibinfo {title} {Localization from {Hilbert} space shattering: {From} theory to physical realizations},\ }\href {https://doi.org/10.1103/PhysRevB.101.174204} {\bibfield  {journal} {\bibinfo  {journal} {Phys. Rev. B}\ }\textbf {\bibinfo {volume} {101}},\ \bibinfo {pages} {174204} (\bibinfo {year} {2020})}\BibitemShut {NoStop}%
\bibitem [{\citenamefont {Zadnik}\ and\ \citenamefont {Fagotti}(2021)}]{zadnik_folded_2021}%
  \BibitemOpen
  \bibfield  {author} {\bibinfo {author} {\bibfnamefont {L.}~\bibnamefont {Zadnik}}\ and\ \bibinfo {author} {\bibfnamefont {M.}~\bibnamefont {Fagotti}},\ }\bibfield  {title} {\bibinfo {title} {The {Folded} {Spin}-1/2 {XXZ} {Model}: {I}. {Diagonalisation}, {Jamming}, and {Ground} {State} {Properties}},\ }\href {https://doi.org/10.21468/SciPostPhysCore.4.2.010} {\bibfield  {journal} {\bibinfo  {journal} {SciPost Phys. Core}\ }\textbf {\bibinfo {volume} {4}},\ \bibinfo {pages} {010} (\bibinfo {year} {2021})}\BibitemShut {NoStop}%
\bibitem [{\citenamefont {Pozsgay}\ \emph {et~al.}(2021)\citenamefont {Pozsgay}, \citenamefont {Gombor}, \citenamefont {Hutsalyuk}, \citenamefont {Jiang}, \citenamefont {Pristy{\'a}k},\ and\ \citenamefont {Vernier}}]{pozsgay2021integrable}%
  \BibitemOpen
  \bibfield  {author} {\bibinfo {author} {\bibfnamefont {B.}~\bibnamefont {Pozsgay}}, \bibinfo {author} {\bibfnamefont {T.}~\bibnamefont {Gombor}}, \bibinfo {author} {\bibfnamefont {A.}~\bibnamefont {Hutsalyuk}}, \bibinfo {author} {\bibfnamefont {Y.}~\bibnamefont {Jiang}}, \bibinfo {author} {\bibfnamefont {L.}~\bibnamefont {Pristy{\'a}k}},\ and\ \bibinfo {author} {\bibfnamefont {E.}~\bibnamefont {Vernier}},\ }\bibfield  {title} {\bibinfo {title} {{Integrable spin chain with Hilbert space fragmentation and solvable real-time dynamics}},\ }\href {https://doi.org/10.1103/PhysRevE.104.044106} {\bibfield  {journal} {\bibinfo  {journal} {Phys. Rev. E}\ }\textbf {\bibinfo {volume} {104}},\ \bibinfo {pages} {044106} (\bibinfo {year} {2021})}\BibitemShut {NoStop}%
\bibitem [{\citenamefont {Moudgalya}\ \emph {et~al.}(2021)\citenamefont {Moudgalya}, \citenamefont {Prem}, \citenamefont {Nandkishore}, \citenamefont {Regnault},\ and\ \citenamefont {Bernevig}}]{moudgalya_thermalization_2021}%
  \BibitemOpen
  \bibfield  {author} {\bibinfo {author} {\bibfnamefont {S.}~\bibnamefont {Moudgalya}}, \bibinfo {author} {\bibfnamefont {A.}~\bibnamefont {Prem}}, \bibinfo {author} {\bibfnamefont {R.}~\bibnamefont {Nandkishore}}, \bibinfo {author} {\bibfnamefont {N.}~\bibnamefont {Regnault}},\ and\ \bibinfo {author} {\bibfnamefont {B.~A.}\ \bibnamefont {Bernevig}},\ }\bibfield  {title} {\bibinfo {title} {Thermalization and its absence within {Krylov} subspaces of a constrained {Hamiltonian}}\ }(\bibinfo {year} {2021})\ pp.\ \bibinfo {pages} {147--209},\ \bibinfo {note} {arXiv:1910.14048 [cond-mat]}\BibitemShut {NoStop}%
\bibitem [{\citenamefont {Maassarani}(1998)}]{maassarani_xxc_1998}%
  \BibitemOpen
  \bibfield  {author} {\bibinfo {author} {\bibfnamefont {Z.}~\bibnamefont {Maassarani}},\ }\bibfield  {title} {\bibinfo {title} {The {XXC} models},\ }\href {https://doi.org/https://doi.org/10.1016/S0375-9601(98)00322-3} {\bibfield  {journal} {\bibinfo  {journal} {Phys. Lett. A}\ }\textbf {\bibinfo {volume} {244}},\ \bibinfo {pages} {160} (\bibinfo {year} {1998})}\BibitemShut {NoStop}%
\bibitem [{\citenamefont {Arnaudon}\ and\ \citenamefont {Maassarani}(1998)}]{arnaudon1998integrable}%
  \BibitemOpen
  \bibfield  {author} {\bibinfo {author} {\bibfnamefont {D.}~\bibnamefont {Arnaudon}}\ and\ \bibinfo {author} {\bibfnamefont {Z.}~\bibnamefont {Maassarani}},\ }\bibfield  {title} {\bibinfo {title} {{Integrable open boundary conditions for XXC models}},\ }\href {https://doi.org/10.1088/1126-6708/1998/10/024} {\bibfield  {journal} {\bibinfo  {journal} {J. High Energy Phys.}\ }\textbf {\bibinfo {volume} {1998}}\bibinfo  {number} { (10)},\ \bibinfo {pages} {024}}\BibitemShut {NoStop}%
\bibitem [{\citenamefont {Zhang}\ and\ \citenamefont {Mussardo}(2022)}]{zhang2022hidden}%
  \BibitemOpen
\bibfield  {number} {  }\bibfield  {author} {\bibinfo {author} {\bibfnamefont {Z.}~\bibnamefont {Zhang}}\ and\ \bibinfo {author} {\bibfnamefont {G.}~\bibnamefont {Mussardo}},\ }\bibfield  {title} {\bibinfo {title} {{Hidden Bethe states in a partially integrable model}},\ }\href {https://doi.org/10.1103/PhysRevB.106.134420} {\bibfield  {journal} {\bibinfo  {journal} {Phys. Rev. B}\ }\textbf {\bibinfo {volume} {106}},\ \bibinfo {pages} {134420} (\bibinfo {year} {2022})}\BibitemShut {NoStop}%
\bibitem [{\citenamefont {Matsui}(2024)}]{Matsui_2024_1}%
  \BibitemOpen
  \bibfield  {author} {\bibinfo {author} {\bibfnamefont {C.}~\bibnamefont {Matsui}},\ }\bibfield  {title} {\bibinfo {title} {{Exactly solvable subspaces of nonintegrable spin chains with boundaries and quasiparticle interactions}},\ }\href {https://doi.org/10.1103/PhysRevB.109.104307} {\bibfield  {journal} {\bibinfo  {journal} {Phys. Rev. B}\ }\textbf {\bibinfo {volume} {109}},\ \bibinfo {pages} {104307} (\bibinfo {year} {2024})}\BibitemShut {NoStop}%
\bibitem [{\citenamefont {Matsui}\ and\ \citenamefont {Tsuji}(2024)}]{matsui2024boundary}%
  \BibitemOpen
  \bibfield  {author} {\bibinfo {author} {\bibfnamefont {C.}~\bibnamefont {Matsui}}\ and\ \bibinfo {author} {\bibfnamefont {N.}~\bibnamefont {Tsuji}},\ }\href@noop {} {\bibinfo {title} {{Boundary dissipative spin chains with partial solvability inherited from system Hamiltonians}}} (\bibinfo {year} {2024}),\ \Eprint {https://arxiv.org/abs/2409.03208} {arXiv:2409.03208 [cond-mat.stat-mech]} \BibitemShut {NoStop}%
\bibitem [{\citenamefont {Laflorencie}\ \emph {et~al.}(2006)\citenamefont {Laflorencie}, \citenamefont {S{\o}rensen}, \citenamefont {Chang},\ and\ \citenamefont {Affleck}}]{laflorencie2006boundary}%
  \BibitemOpen
  \bibfield  {author} {\bibinfo {author} {\bibfnamefont {N.}~\bibnamefont {Laflorencie}}, \bibinfo {author} {\bibfnamefont {E.~S.}\ \bibnamefont {S{\o}rensen}}, \bibinfo {author} {\bibfnamefont {M.-S.}\ \bibnamefont {Chang}},\ and\ \bibinfo {author} {\bibfnamefont {I.}~\bibnamefont {Affleck}},\ }\bibfield  {title} {\bibinfo {title} {Boundary effects in the critical scaling of entanglement entropy in {1D} systems},\ }\href {https://doi.org/10.1103/PhysRevLett.96.100603} {\bibfield  {journal} {\bibinfo  {journal} {Phys. Rev. Lett.}\ }\textbf {\bibinfo {volume} {96}},\ \bibinfo {pages} {100603} (\bibinfo {year} {2006})}\BibitemShut {NoStop}%
\bibitem [{\citenamefont {D'Emidio}\ \emph {et~al.}(2015)\citenamefont {D'Emidio}, \citenamefont {Block},\ and\ \citenamefont {Kaul}}]{demidio_2015}%
  \BibitemOpen
  \bibfield  {author} {\bibinfo {author} {\bibfnamefont {J.}~\bibnamefont {D'Emidio}}, \bibinfo {author} {\bibfnamefont {M.~S.}\ \bibnamefont {Block}},\ and\ \bibinfo {author} {\bibfnamefont {R.~K.}\ \bibnamefont {Kaul}},\ }\bibfield  {title} {\bibinfo {title} {R\'enyi entanglement entropy of critical $\mathrm{SU}(n)$ spin chains},\ }\href {https://doi.org/10.1103/PhysRevB.92.054411} {\bibfield  {journal} {\bibinfo  {journal} {Phys. Rev. B}\ }\textbf {\bibinfo {volume} {92}},\ \bibinfo {pages} {054411} (\bibinfo {year} {2015})}\BibitemShut {NoStop}%
\bibitem [{\citenamefont {Kim}\ \emph {et~al.}(2016)\citenamefont {Kim}, \citenamefont {Katsura}, \citenamefont {Trivedi},\ and\ \citenamefont {Han}}]{kim_2016}%
  \BibitemOpen
  \bibfield  {author} {\bibinfo {author} {\bibfnamefont {P.}~\bibnamefont {Kim}}, \bibinfo {author} {\bibfnamefont {H.}~\bibnamefont {Katsura}}, \bibinfo {author} {\bibfnamefont {N.}~\bibnamefont {Trivedi}},\ and\ \bibinfo {author} {\bibfnamefont {J.~H.}\ \bibnamefont {Han}},\ }\bibfield  {title} {\bibinfo {title} {Entanglement and corner hamiltonian spectra of integrable open spin chains},\ }\href {https://doi.org/10.1103/PhysRevB.94.195110} {\bibfield  {journal} {\bibinfo  {journal} {Phys. Rev. B}\ }\textbf {\bibinfo {volume} {94}},\ \bibinfo {pages} {195110} (\bibinfo {year} {2016})}\BibitemShut {NoStop}%
\bibitem [{\citenamefont {Calabrese}\ and\ \citenamefont {Cardy}(2004)}]{calabrese_entanglement_2004}%
  \BibitemOpen
  \bibfield  {author} {\bibinfo {author} {\bibfnamefont {P.}~\bibnamefont {Calabrese}}\ and\ \bibinfo {author} {\bibfnamefont {J.}~\bibnamefont {Cardy}},\ }\bibfield  {title} {\bibinfo {title} {{Entanglement Entropy and Quantum Field Theory}},\ }\href {https://doi.org/10.1088/1742-5468/2004/06/P06002} {\bibfield  {journal} {\bibinfo  {journal} {J. Stat. Mech.: Theor. Exp.}\ }\textbf {\bibinfo {volume} {2004}},\ \bibinfo {pages} {P06002} (\bibinfo {year} {2004})}\BibitemShut {NoStop}%
\bibitem [{\citenamefont {Nomura}\ and\ \citenamefont {Okamoto}(1994)}]{K_Nomura_1994}%
  \BibitemOpen
  \bibfield  {author} {\bibinfo {author} {\bibfnamefont {K.}~\bibnamefont {Nomura}}\ and\ \bibinfo {author} {\bibfnamefont {K.}~\bibnamefont {Okamoto}},\ }\bibfield  {title} {\bibinfo {title} {{Critical properties of $S=1/2$ antiferromagnetic {XXZ} chain with next-nearest-neighbour interactions}},\ }\href {https://doi.org/10.1088/0305-4470/27/17/012} {\bibfield  {journal} {\bibinfo  {journal} {J. Phys. A: Math. Gen.}\ }\textbf {\bibinfo {volume} {27}},\ \bibinfo {pages} {5773} (\bibinfo {year} {1994})}\BibitemShut {NoStop}%
\bibitem [{\citenamefont {Ueda}\ and\ \citenamefont {Oshikawa}(2021)}]{ueda_resolving_2021}%
  \BibitemOpen
  \bibfield  {author} {\bibinfo {author} {\bibfnamefont {A.}~\bibnamefont {Ueda}}\ and\ \bibinfo {author} {\bibfnamefont {M.}~\bibnamefont {Oshikawa}},\ }\bibfield  {title} {\bibinfo {title} {Resolving the {Berezinskii}-{Kosterlitz}-{Thouless} transition in the two-dimensional {XY} model with tensor-network-based level spectroscopy},\ }\href {https://doi.org/10.1103/PhysRevB.104.165132} {\bibfield  {journal} {\bibinfo  {journal} {Phys. Rev. B}\ }\textbf {\bibinfo {volume} {104}},\ \bibinfo {pages} {165132} (\bibinfo {year} {2021})}\BibitemShut {NoStop}%
\bibitem [{\citenamefont {Bernien}\ \emph {et~al.}(2017)\citenamefont {Bernien}, \citenamefont {Schwartz}, \citenamefont {Keesling}, \citenamefont {Levine}, \citenamefont {Omran}, \citenamefont {Pichler}, \citenamefont {Choi}, \citenamefont {Zibrov}, \citenamefont {Endres}, \citenamefont {Greiner}, \citenamefont {Vuletić},\ and\ \citenamefont {Lukin}}]{bernien_probing_2017}%
  \BibitemOpen
  \bibfield  {author} {\bibinfo {author} {\bibfnamefont {H.}~\bibnamefont {Bernien}}, \bibinfo {author} {\bibfnamefont {S.}~\bibnamefont {Schwartz}}, \bibinfo {author} {\bibfnamefont {A.}~\bibnamefont {Keesling}}, \bibinfo {author} {\bibfnamefont {H.}~\bibnamefont {Levine}}, \bibinfo {author} {\bibfnamefont {A.}~\bibnamefont {Omran}}, \bibinfo {author} {\bibfnamefont {H.}~\bibnamefont {Pichler}}, \bibinfo {author} {\bibfnamefont {S.}~\bibnamefont {Choi}}, \bibinfo {author} {\bibfnamefont {A.~S.}\ \bibnamefont {Zibrov}}, \bibinfo {author} {\bibfnamefont {M.}~\bibnamefont {Endres}}, \bibinfo {author} {\bibfnamefont {M.}~\bibnamefont {Greiner}}, \bibinfo {author} {\bibfnamefont {V.}~\bibnamefont {Vuletić}},\ and\ \bibinfo {author} {\bibfnamefont {M.~D.}\ \bibnamefont {Lukin}},\ }\bibfield  {title} {\bibinfo {title} {Probing many-body dynamics on a 51-atom quantum simulator},\ }\href {https://doi.org/10.1038/nature24622} {\bibfield  {journal} {\bibinfo  {journal} {Nature}\ }\textbf {\bibinfo {volume} {551}},\
  \bibinfo {pages} {579} (\bibinfo {year} {2017})}\BibitemShut {NoStop}%
\bibitem [{\citenamefont {Su}\ \emph {et~al.}(2023)\citenamefont {Su}, \citenamefont {Sun}, \citenamefont {Hudomal}, \citenamefont {Desaules}, \citenamefont {Zhou}, \citenamefont {Yang}, \citenamefont {Halimeh}, \citenamefont {Yuan}, \citenamefont {Papi\ifmmode~\acute{c}\else \'{c}\fi{}},\ and\ \citenamefont {Pan}}]{su2023observation}%
  \BibitemOpen
  \bibfield  {author} {\bibinfo {author} {\bibfnamefont {G.-X.}\ \bibnamefont {Su}}, \bibinfo {author} {\bibfnamefont {H.}~\bibnamefont {Sun}}, \bibinfo {author} {\bibfnamefont {A.}~\bibnamefont {Hudomal}}, \bibinfo {author} {\bibfnamefont {J.-Y.}\ \bibnamefont {Desaules}}, \bibinfo {author} {\bibfnamefont {Z.-Y.}\ \bibnamefont {Zhou}}, \bibinfo {author} {\bibfnamefont {B.}~\bibnamefont {Yang}}, \bibinfo {author} {\bibfnamefont {J.~C.}\ \bibnamefont {Halimeh}}, \bibinfo {author} {\bibfnamefont {Z.-S.}\ \bibnamefont {Yuan}}, \bibinfo {author} {\bibfnamefont {Z.}~\bibnamefont {Papi\ifmmode~\acute{c}\else \'{c}\fi{}}},\ and\ \bibinfo {author} {\bibfnamefont {J.-W.}\ \bibnamefont {Pan}},\ }\bibfield  {title} {\bibinfo {title} {{Observation of many-body scarring in a Bose-Hubbard quantum simulator}},\ }\href {https://doi.org/10.1103/PhysRevResearch.5.023010} {\bibfield  {journal} {\bibinfo  {journal} {Phys. Rev. Res.}\ }\textbf {\bibinfo {volume} {5}},\ \bibinfo {pages} {023010} (\bibinfo {year}
  {2023})}\BibitemShut {NoStop}%
\bibitem [{Note6()}]{Note6}%
  \BibitemOpen
  \bibinfo {note} {If the particle number is not fixed, the ground state becomes $L$-fold degenerate due to the emergent SU(2) symmetry in the effective XXZ model with $|J_x|=|J_z|$. The $\eta $-operators $\eta $ and $\eta ^\dagger $ relate each eigenmode, but they are not localized at the edges.}\BibitemShut {Stop}%
\bibitem [{\citenamefont {Scaffidi}\ \emph {et~al.}(2017)\citenamefont {Scaffidi}, \citenamefont {Parker},\ and\ \citenamefont {Vasseur}}]{gaplessSPT}%
  \BibitemOpen
  \bibfield  {author} {\bibinfo {author} {\bibfnamefont {T.}~\bibnamefont {Scaffidi}}, \bibinfo {author} {\bibfnamefont {D.~E.}\ \bibnamefont {Parker}},\ and\ \bibinfo {author} {\bibfnamefont {R.}~\bibnamefont {Vasseur}},\ }\bibfield  {title} {\bibinfo {title} {Gapless symmetry-protected topological order},\ }\href {https://doi.org/10.1103/PhysRevX.7.041048} {\bibfield  {journal} {\bibinfo  {journal} {Phys. Rev. X}\ }\textbf {\bibinfo {volume} {7}},\ \bibinfo {pages} {041048} (\bibinfo {year} {2017})}\BibitemShut {NoStop}%
\bibitem [{\citenamefont {Yoshida}\ \emph {et~al.}(2024)\citenamefont {Yoshida}, \citenamefont {Heinsdorf},\ and\ \citenamefont {Katsura}}]{yoshida_edge_2024_zenodo}%
  \BibitemOpen
  \bibfield  {author} {\bibinfo {author} {\bibfnamefont {H.}~\bibnamefont {Yoshida}}, \bibinfo {author} {\bibfnamefont {N.}~\bibnamefont {Heinsdorf}},\ and\ \bibinfo {author} {\bibfnamefont {H.}~\bibnamefont {Katsura}},\ }\bibfield  {title} {\bibinfo {title} {{Data of ``Edge-Edge Correlations without Edge-States: {$\eta$}-clustering State as Ground State of the Extended Attractive SU(3) Hubbard Chain''}},\ }\bibfield  {journal} {\bibinfo  {journal} {Zenodo}\ }\href {https://doi.org/10.5281/zenodo.14682000} {10.5281/zenodo.14682000} (\bibinfo {year} {2024})\BibitemShut {NoStop}%
\bibitem [{\citenamefont {White}(1993)}]{white1992density}%
  \BibitemOpen
  \bibfield  {author} {\bibinfo {author} {\bibfnamefont {S.~R.}\ \bibnamefont {White}},\ }\bibfield  {title} {\bibinfo {title} {Density-matrix algorithms for quantum renormalization groups},\ }\href {https://doi.org/10.1103/PhysRevB.48.10345} {\bibfield  {journal} {\bibinfo  {journal} {Phys. Rev. B}\ }\textbf {\bibinfo {volume} {48}},\ \bibinfo {pages} {10345} (\bibinfo {year} {1993})}\BibitemShut {NoStop}%
\bibitem [{\citenamefont {Hauschild}\ and\ \citenamefont {Pollmann}(2018)}]{hauschild2018efficient}%
  \BibitemOpen
  \bibfield  {author} {\bibinfo {author} {\bibfnamefont {J.}~\bibnamefont {Hauschild}}\ and\ \bibinfo {author} {\bibfnamefont {F.}~\bibnamefont {Pollmann}},\ }\bibfield  {title} {\bibinfo {title} {{Efficient numerical simulations with Tensor Networks: Tensor Network Python (TeNPy)}},\ }\href {https://doi.org/10.21468/SciPostPhysLectNotes.5} {\bibfield  {journal} {\bibinfo  {journal} {SciPost Phys. Lect. Notes}\ ,\ \bibinfo {pages} {5}} (\bibinfo {year} {2018})}\BibitemShut {NoStop}%
\bibitem [{\citenamefont {Schollw{\"o}ck}(2011)}]{schollwock2011density}%
  \BibitemOpen
  \bibfield  {author} {\bibinfo {author} {\bibfnamefont {U.}~\bibnamefont {Schollw{\"o}ck}},\ }\bibfield  {title} {\bibinfo {title} {The density-matrix renormalization group in the age of matrix product states},\ }\href {https://doi.org/10.1016/j.aop.2010.09.012} {\bibfield  {journal} {\bibinfo  {journal} {Ann. Phys.}\ }\textbf {\bibinfo {volume} {326}},\ \bibinfo {pages} {96} (\bibinfo {year} {2011})}\BibitemShut {NoStop}%
\end{thebibliography}%

\end{document}